%% file: main.tex
\documentclass[fleqn,10pt]{wlscirep}

\usepackage[utf8]{inputenc}
\usepackage[T1]{fontenc}

\graphicspath{{./figs/main/}}
\usepackage{paralist}
\usepackage{enumerate}
\usepackage{multirow}
\usepackage{mathtools}
\usepackage{graphicx}
\usepackage{float}
\usepackage{bm}

\usepackage{caption}
\usepackage{subcaption}
\newcommand{\SecRef}[1]{Section~\ref{#1}}
\newcommand{\ManuSecRef}[1]{Manuscript Section ``\textit{#1}''}
\newcommand{\eqRef}[1]{equation~\eqref{#1}}
\newcommand{\EqRef}[1]{Equation~\eqref{#1}}
\newcommand{\figRef}[1]{figure~\ref{#1}}
\newcommand{\FigRef}[1]{Figure~\ref{#1}}
\newcommand{\tabRef}[1]{table~\ref{#1}}
\newcommand{\TabRef}[1]{Table~\ref{#1}}

\newcommand{\AlgRef}[1]{Algorithm~\ref{#1}}
\newcommand{\refCite}[1]{reference\phantom{ }\cite{#1}}

\newcommand\tinybullet{$\vcenter{\hbox{\tiny$\bullet$}}\;$}

\usepackage{xcolor}
\newcommand{\red}[1]{#1}       
\newcommand{\blue}[1]{#1}     
\newcommand{\violet}[1]{#1} 

\usepackage[noend]{algpseudocode}
\usepackage[ruled,vlined]{algorithm2e}
\SetKwInput{Input}{Input}                
\SetKwInput{Output}{Output}              
\newcommand{\algDefine}{\textbf{define} }

\newcommand{\algBreak}{\textbf{break} }
\newcommand{\algAnd}{{ \normalfont\textbf{and} }}

\renewcommand*{\thesection}{}
\renewcommand*{\thesubsection}{\arabic{subsection}}
\renewcommand*{\thesubsubsection}{\thesubsection.\arabic{subsubsection}}

\makeatletter
\newcommand*{\addFileDependency}[1]{
	\typeout{(#1)}
	\@addtofilelist{#1}
	\IfFileExists{#1}{}{\typeout{No file #1.}}
}
\makeatother

\newcommand*{\myexternaldocument}[1]{%
	\externaldocument{#1}%
	\addFileDependency{#1.tex}%
	\addFileDependency{#1.aux}%
}
\myexternaldocument{suppl}

\title{Adaptive physics-informed neural operator for coarse-grained non-equilibrium flows}

\author[1]{Ivan Zanardi}
\author[1]{Simone Venturi}
\author[1,*]{Marco Panesi}
\affil[1]{Center for Hypersonics and Entry Systems Studies, Department of Aerospace Engineering, University of Illinois Urbana-Champaign, Urbana, IL 61801, USA}

\affil[*]{mpanesi@illinois.edu}


\begin{abstract}
	This work proposes a new machine learning (ML)-based paradigm aiming to enhance the computational efficiency of non-equilibrium reacting flow simulations while ensuring compliance with the underlying physics. The framework combines dimensionality reduction and neural operators through a hierarchical and adaptive deep learning strategy to learn the solution of multi-scale coarse-grained governing equations for chemical kinetics. The proposed surrogate's architecture is structured as a tree, with leaf nodes representing separate neural operator blocks where physics is embedded in the form of multiple soft and hard constraints. The hierarchical attribute has two advantages: i) It allows the simplification of the training phase via transfer learning, starting from the slowest temporal scales; ii) It accelerates the prediction step by enabling adaptivity as the surrogate's evaluation is limited to the necessary leaf nodes based on the local degree of non-equilibrium of the gas. The model is applied to the study of chemical kinetics relevant for application to hypersonic flight, and it is tested here on pure oxygen gas mixtures. \violet{In 0-$\mathrm{D}$ scenarios, the proposed ML framework can adaptively predict the dynamics of almost thirty species with a maximum relative error of 4.5\% for a wide range of initial conditions. Furthermore, when employed in 1-$\mathrm{D}$ shock simulations, the approach shows accuracy ranging from 1\% to 4.5\% and a speedup of one order of magnitude compared to conventional implicit schemes employed in an operator-splitting integration framework. Given the results presented in the paper, this work lays the foundation for constructing an efficient ML-based surrogate coupled with reactive Navier-Stokes solvers for accurately characterizing non-equilibrium phenomena in multi-dimensional computational fluid dynamics simulations.}
\end{abstract}

\begin{document}

\flushbottom
\maketitle
\thispagestyle{empty}

\section*{Introduction}\label{sec:intro}
Accurate modeling of non-equilibrium reacting flows is critical in many engineering and science disciplines, \textit{e.g.}, designing hypersonic vehicles for space exploration~\cite{Gnoffo_ARFM_1999,Johnston_PRF_2018} or material processing and manufacturing with low-temperature plasmas~\cite{Harpale_JCP_2015,Harpale_PR_2016}. The need for describing and understanding these flows has led to the development of increasingly large and sophisticated mathematical models~\cite{Panesi_JCP_2013,Panesi_PR_2014,Munafo_PoP_2013,Kustova_AIP_2019}, describing multiple physical phenomena characterized by a broad spectrum of spatio-temporal scales.
\par
The most physically consistent approach to model non-equilibrium flows relies on the direct numerical solution of the master equation~\cite{Nagnibeda_Book_2009,Panesi_JTHT_2011,Panesi_JCP_2013,Panesi_PR_2014,Macdonald_PRF_2016,Capitelli_Book_2016,Macdonald_JFC_2020}, whereby all the relevant spatial and temporal scales resulting from chemical and radiative processes are accounted for. Indeed, the availability of quantum state-to-state (StS) chemistry models based on \textit{ab initio} theories~\cite{Wang_JCP_2003,Esposito_CP_2006,Galvao_JPC_2009,Jaffe_AIAA_2009,Venturi_JCPA_2020} enables unprecedented levels of physical accuracy~\cite{Panesi_JCP_2013,Panesi_PR_2014,Munafo_PoP_2013,Kustova_AIP_2019}, crucial for modeling flows typified by a significant degree of non-equilibrium. However, the exponentially large number of degrees of freedom (\textit{i.e.}, molecules' and atoms' energy levels) and the numerical restrictions (stiffness) associated with the derived system of equations make these models impracticable in large-scale multi-dimensional computational fluid dynamics (CFD) problems. To overcome these difficulties, crude ``engineering'' non-equilibrium models~\cite{Hammerling_PoF_1959,Knab_JTHT_1995}, referred to as multi-temperature (MT) models, have been developed over the years, often assembled without any rigorous derivation from fundamental kinetic equations nor consideration for physical principles and constraints. Given their interpolative nature, these cannot be used to perform predictions outside their development range.
\par
This work targets the numerical challenges in solving such computationally intense systems of equations by surrogating the thermochemical processes characterizing non-equilibrium phenomena that conventional techniques cannot address. Surrogate and reduced-order models~\cite{Zhu_JComP_2019,Haghighat_CMAME_2021,Sun_CMAME_2020,Choi_JComP_2021,You_CMAME_2021,Mai_IJUQ_2016} can be designed and constructed by employing various techniques, such as projection-based methods~\cite{Rozza_ACME_2007,Benner_SIAM_2015,Amsallem_IJNME_2012,Huang_JComP_2022,Swischuk_CF_2019,Choi_SIAM_2019,Carlberg_IJNME_2011}, data-fit interpolation and regression~\cite{Forrester_Book_2008}, and machine learning (ML)-based models~\cite{Xu_CMAME_2020,Kim_JComP_2022}. A recent application of surrogates for hypersonics has been published by Ozbenli \textit{et al.}~\cite{Ozbenli_AIAA_2020}, who trained a feed-forward neural network (FNN) to learn a given set of the master equations' solution functions for a specific non-equilibrium model~\cite{Colonna_JTHT_2006}. Their ML framework showed a great computational speed-up compared to numerical integrators, with generalization performances left unclear. Similarly, Campoli \textit{et al.}~\cite{Campoli_Math_2022} explored different ML algorithms to regress the source terms of the ODEs system modeling the thermochemical relaxation processes. A coupling between a conventional integrator and the ML regressor was attempted, and speed-up performances were analyzed. They also tried to infer the solution of Euler's equations for a single one-dimensional reacting shock flow scenario by leveraging a deep neural network (DNN). Scherding and coworkers~\cite{Scherding_arXiv_2022} developed a lower-dimensional surrogate to compute the thermochemical properties of the gas mixture to be used in place of any high-dimensional look-up non-equilibrium thermodynamic library. However, despite the considerable speed-up performances and encouraging perspective, they did it only for steady-state solutions, targeting specific flow conditions and considering only chemical and not thermal non-equilibrium. The above-mentioned frameworks lack generalization performances and do not impose physical constraints during the surrogate construction, making them less suitable for CFD simulations. Instead, the present study aims to provide a prototyping tool that can replace the master equations with a surrogate that preserves the original's essential properties and physical constraints while being orders of magnitude faster and able to cover an extensive range of physical conditions. The present work augments the framework introduced by Zanardi \textit{et al.}~\cite{Zanardi_AIAA_2022}, and it introduces a new machine learning-based method for solving non-equilibrium flows by combining:
\begin{enumerate}[i.]
	\item Coarse-graining, \textit{i.e.}, a reduced order modeling (ROM) technique that extracts meaningful physics from the master equations~\cite{Panesi_JTHT_2011,Panesi_Lani_PofF_2013,Munafo_PR_2014,Munafo_PF_2015,Liu_JCP_2015}, in general, by leveraging unsupervised learning adaptation to seek the optimal grouping configuration~\cite{Sahai_JCP_2017}. The so-derived reduced system of equations models the dynamics of groups of states, addressing the high-dimensionality problem characterizing the StS models.
	\item Neural operators, \textit{i.e.}, a ML-based surrogate that approximates the integral solution operator of a family of partial differential equations (PDEs) to bypass conventional numerical integration~\cite{Kovachki_arXiv_2021}.
\end{enumerate}
\par
\textbf{Coarse-graining}. Constructing a surrogate for high-fidelity quantum-state-specific chemistry models to describe non-equilibrium phenomena is not a simple task as they rely on the solution of an overwhelmingly large number of differential equations (order of 10$^5$)~\cite{Panesi_JCP_2013}. More importantly, the mathematical closure of these equations requires the determination of a sizeable kinetic database that often cannot be computed owing to many processes (order of 10$^{16}$) to be considered. Therefore, performing first a physics-preserving dimensionality reduction is of paramount importance. To this end, nonlinear manifold learning techniques such as autoencoders~\cite{Kingma_arXiv_2013}, diffusion maps~\cite{Coifman_PNAS_2005}, or kernel PCA~\cite{Scholkopf_NC_1998} could be employed. Recently, Oommen \textit{et al.}~\cite{Oommen_arXiv_2022} proposed learning high-dimensional complex dynamics by combining neural operators and autoencoders. Their application first reduced the problem's dimensionality by training a convolutional autoencoder and then learned the low-dimensional dynamics lying in the latent space using a deep neural operator. However, although powerful in applications requiring dimensionality reduction, autoencoders lack physical interpretability and introduce spurious correlations, not necessarily guaranteeing a discrete separation of temporal scales. To overcome these limitations, our approach relies on a class of physics-based reduced-order coarse-grained (CG) models~\cite{Merchant_ARCC_2011,Kmiecik_CR_2016,Boniecki_NAR_2015}. In chemical kinetics, coarse-grained modeling has extensively been used to describe non-equilibrium phenomena of atomic and molecular species~\cite{Heritier_JCP_2014, Esposito_CP_2000, Liu_JCP_2015, Sahai_JCP_2017, Venturi_JCP_2020, Sharma_PR_2020}. The central idea in the proposed CG model is to combine the solution of the coarse-grained dynamics with the partial equilibration of the underlying microscopic structure. The concept of partial equilibrium suggests applying the \textit{maximum entropy principles} (MEP) to reconstruct the unresolved scales or physics. This choice is of paramount importance, as it ensures the physical consistency of the model by enforcing the \textit{principle of detailed balance} and ensuring the positivity and boundness of the distribution function.
\par
\textbf{Neural operators}. The second basis of the proposed methodology aims to address the stiffness associated with thermochemical processes, characterized by a broad spectrum of temporal scales, ranging from the flow time scales to time scales that are orders of magnitude smaller. This work uses DNNs to infer the generalized solution of the governing equations to bypass the conventional numerical integration. In literature, a series of new ML-based paradigms for speeding up the numerical simulation of partial differential equations~\cite{Jagtap_CCP_2020,Bar_arXiv_2019,Bhatnagar_CM_2019,Zhu_JComP_2018,Sirignano_JComP_2018,Duvall_arXiv_2021,Gao_JComP_2021} have been proposed over the past few years. In particular, this work leverages the family of neural operators~\cite{Kissas_arXiv_2022,Lu_CMAME_2022,Li_arXiv_2020,Li_arXiv_2020_Fourier,Kovachki_arXiv_2021,You_JComP_2022}, DNN-based surrogates designed to learn or discover solution operators defined by the mapping between inputs of a dynamical system, such as initial or boundary conditions (ICs/BCs), and its state. We employ a parametric-based approach to operator learning, introduced first by Chen \textit{et al.}~\cite{Chen_IEEE_1995} and then recently extended by Lu \textit{et al.}~\cite{Lu_NMI_2021} In their work, Lu and coworkers introduced DeepONet, a novel network architecture that effectively approximates the solution operator of linear and nonlinear parametric PDEs. DeepONets have found applications in various fields of physics~\cite{Ranade_arXiv_2021,Sharma_CRxiv_2022}, including hypersonics with the work of Mao \textit{et al.}~\cite{Mao_JComP_2021}, who approximated the fluid flow evolution and concentration profiles downstream of a normal shock with a DeepONet-based surrogate. Although Mao \textit{et al.}'s work is significant for the scientific community, it relies on a simple physical model that cannot correctly represent the non-equilibrium distribution of internal energy states, which is crucial for the current study. Additionally, their approach lacks physics constraints during the design and training phase of the model, such as physics-informed (PI) machine learning methodologies employed in this work, commonly known as PINNs~\cite{Raissi_JComP_2019,Karniadakis_NRP_2021,Shukla_JComP_2021,Wang_PRF_2017,Mao_CMAME_2020,Jagtap_CMAME_2020}. These techniques impose constraints by penalizing deviations from governing equations, enhancing the model's generalization performance. This new class of machine learning models, called physics-informed deep neural operator (PI-DeepONet)~\cite{Wang_SA_2021,WangPerdikaris_arxiv_2021,Wang_JSC_2022,Goswami_arXiv_2022}, which combines physics-informed techniques with the DeepONet architecture, was initially introduced by Wang \textit{et al.}~\cite{Wang_SA_2021} and successfully applied to construct surrogate solution operators for various partial differential equations (PDEs), demonstrating excellent results.
\begin{figure}[!t]
    \centering
	\includegraphics[width=0.8\textwidth]{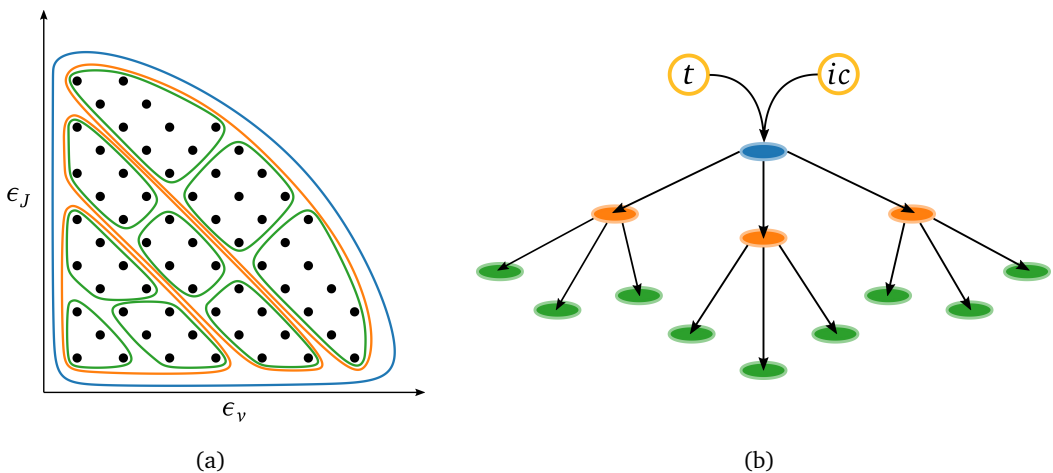}
	\caption{\textit{Schematics of the proposed approach.} Combining coarse-graining (a) and hierarchical DeepONets (b). (a) Reduced order modeling technique based on clustering the species' quantum energy states (schematized as black dots and as functions of vibrational, $\epsilon_v$, and rotational energy, $\epsilon_J$) into macroscopic bins. In the figure, three different levels of hierarchical clustering are shown. (b) Tree visualization of the hierarchical deep learning framework, where the leaf nodes correspond to separate DeepONets (one for each macroscopic bin), which take as inputs the initial conditions, $ic$, and time, $t$.}
    \label{fig:intro.approach}
\end{figure}
\par
\textbf{Proposed approach}. The combined use of coarse-graining and neural operators is of primary importance. On the one hand, the mere application of neural operators does not resolve the high-dimensionality problem, as it is not straightforward to design and train an efficient surrogate for thousands of coupled differential equations. On the other hand, dimensionality reduction does not solve the issues with integration, as small steps are still needed to stably integrate the reduced system of equations. For these reasons, the proposed framework is characterized by a novel physics-inspired architecture based on a hierarchy of DeepONets used to learn the solution operator for multiple coarse-grained configurations to resolve different scales of the phenomena considered. The CG surrogate herein proposed, referred to as CG-DeepONet throughout the rest of the paper, is constructed by training each scale sequentially and employing transfer learning between them. In this sense, our framework is in line with recent operator learning techniques for multi-scale systems~\cite{Liu_PTRS_2022,Migus_arXiv_2022,Liu_arXiv_2022,Liu_arXiv_2021,Lin_JCP_2021,Lutjens_arXiv_2022}. Among the latest ones, Liu \textit{et al.}~\cite{Liu_PTRS_2022} proposed a promising hierarchical time-stepper approach for solving the system dynamics. In their approach, they trained multiple neural networks to capture different timescales of the physical phenomenon by varying the integration step. We also recall the work of Migus \textit{et al.}~\cite{Migus_arXiv_2022}, who designed a multi-scale architecture based on multi-pole graph neural operators (MGNO) by embedding multi-resolution iterative methods~\cite{Jaysaval_GJI_2016}. Liu and coworkers~\cite{Liu_arXiv_2022} drew inspiration from hierarchical matrix methods to develop their multi-scale hierarchical transformer. Furthermore, Liu and Cai~\cite{Liu_arXiv_2021} integrated multi-scale deep neural networks (MscaleDNNs)~\cite{Ziqi_CCP_2020} within the DeepONet architecture. These innovative approaches open up new possibilities for more accurate and efficient modeling of multi-scale complex systems, and the paradigm proposed in this work builds upon these advancements. Indeed, our framework allows the development of a parsimonious and autonomous tool that can quickly deliver the optimal thermochemical representation of the gas given initial conditions and time instant by adaptively choosing the most efficient and physically accurate grouping resolution. The need for adaptation is a direct consequence of different physical scenarios arising in multidimensional numerical simulations, ranging from equilibrium or near-equilibrium to strong non-equilibrium conditions. A controller-acting surrogate, identified as Neq-DeepONet in the remainder of this paper, is responsible for the model adaption to the local flow conditions. In this sense, our framework can be viewed as a multi-fidelity composition of DeepONets and shares analogies with some recent works on the topic~\cite{Thakur_arXiv_2022,Howard_arXiv_2022,Lu_PRR_2022}. However, the novelty of our approach stems from the definition of such a composition based on the maximum-entropy coarse-grained modeling, which is consistent with the underlying physics.
\par
\violet{\textbf{Physics-informed attributes of the surrogate}.
In this paragraphs, we highlight the physics-informed features of the proposed approach, which take the form of either soft or hard constraints imposed on the surrogate:
\begin{enumerate}[i.]
	\item \textit{Dimensionality reduction in the state space}\\
	In addition to the dimensionality reduction in the space of the initial conditions automatically carried out by the DeepONet based on the scenarios provided during training~\cite{Venturi_CMAME_2023}, a physics-based reduction is performed in the state space (\textit{i.e.}, in the space of the discrete energy states) by grouping states that are likely to be found in local equilibrium~\cite{Sahai_JCP_2017,Venturi_JCP_2020}. Only briefly introduced above, such a coarse-graining approach will be detailed in \SecRef{sssec:physics.CG}.
	\item \textit{Physics-consistent architecture components}\\
	A Boltzmann transformation layer is built into the surrogate to enforce the equilibrium distributions between states in the same group, as explained in \SecRef{sssec:ml.architecture}.
	\item \textit{Interpretable prior distributions for the network parameters}\\
	As discussed in \SecRef{sssec:ml.architecture}, the addition of Boltzmann layers allows the imposition of prior distributions for the network parameters that, when propagated to the state populations (\textit{e.g.}, mass fractions), produce equilibrium distributions between distinct groups of states. Therefore, such priors can provide physically consistent solutions even for un-trained surrogates.
	\item \textit{Physics-informed loss function}\\
	The framework employs a physics-informed loss as a soft constraint, which biases the surrogate predictions towards physically consistent solutions. In particular, the employed hybrid strategy, described in \SecRef{sssec:train}, combines data from high-fidelity simulations (or experiments) to anchor the solution to frequent or reproducible real-world scenarios and the residual of the governing laws to ensure generalizability to different unseen physical conditions.
	\item \textit{Hierarchical architecture and transfer learning}\\
	The training strategy involves sequential fine-tuning transfer learning between different temporal scales, explained in \SecRef{sssec:train}. On the one hand, this approach allows for partially preserving the learned physics. On the other hand, it enables surrogate adaptation and knowledge transfer from one temporal scale to another, speeding up the training process of the entire network.
	\item \textit{Physics-driven online pruning at the prediction phase}\\
	As detailed in \SecRef{ssec:ml.adapt_tech}, an additional (controller-acting) surrogate learns the dynamics of a physically-relevant non-equilibrium control variable, determining the minimum resolution level required to accurately describe the system dynamics while avoiding explicitly computing unnecessary fine scales. During the prediction phase, this additional surrogate is responsible for selecting which component of the overall architecture needs to be queried.
\end{enumerate}}
\par
The paper is structured as follows. First, in \SecRef{ssec:physics}, the basic framework and derivation of the thermochemical non-equilibrium model are provided, along with the details of the one-dimensional numerical experiment conducted in this work. Next, in Sections \ref{ssec:ml} and \ref{ssec:ml.adapt_tech}, the proposed ML framework and the developed adaptive technique are described, respectively. In the ``Results'' section, the accuracy and performance of the surrogate with and without adaptive inference are illustrated and discussed in detail for both 0-$\mathrm{D}$ and 1-$\mathrm{D}$ test case scenarios. Finally, in the ``Conclusions'' section, final remarks are presented along with possibilities for future work. Additional information can be found in the Supplementary Information for interested readers.

\section*{Methods}\label{sec:methods}
\subsection{Physical modeling}\label{ssec:physics}
Modeling of chemically reacting flows relies on the solution of Navier-Stokes equations complemented by additional conservation equations accounting for changes in the chemical composition and non-equilibrium relaxation of the energy modes. This extra set of equations often represents a computational burden that makes reacting non-equilibrium flows hard to solve. An extensive discussion on non-equilibrium modeling can be found in \refCite{Liu_JCP_2015}.
\par
The most general way to express the extra set of governing equations is
\begin{equation}\label{eq:gov_eq}
	\frac{D}{Dt} \left(\rho_i e^m_i\right) + \nabla \cdot \mathcal{J}^m_i = \Omega^m_i\hspace{2mm},
\end{equation}
where $\rho_i$ and $e_i$ indicate the mass density and the internal energy of the $i$-th pseudo-species (\textit{i.e.}, a particular species' internal degree of freedom treated as a state variable). Additionally, $m$ denotes the moment order (0, 1, 2, \textit{etc.}), $\Omega^m_i$ the reactive source terms, $D/Dt$ the Lagrangian derivative, and $\mathcal{J}_i^m$ the dissipative/diffusion terms. Depending on the assumptions made in the definition of the chemical species indicated by $i$, three different models can be identified:
\begin{enumerate}[i.]
	\item If $i$ refers to a particular energy state, $\epsilon_i$ (\textit{i.e.}, rovibronic $i=(el,v,J)$), the approach is called state-to-state (StS) master equations~\cite{Panesi_JCP_2013,Panesi_PR_2014}. In this case, $m$ is set to 0.
	\item If $\rho_i$ indicates the density of a group of states, the approach is named coarse-grained (CG) modeling or coarse-grained master equations (CGME)~\cite{Panesi_Lani_PofF_2013,Liu_JCP_2015,Munafo_EPJ_2012,Sahai_JCP_2017,Venturi_JCP_2020,Sharma_PR_2020,Macdonald_I_JCP_2018,Macdonald_II_JCP_2018}. In this case, the conservation equations for mass, momentum, and energy are complemented by additional equations (\textit{i.e.}, $m=0$ and/or $m=1$) to model chemical composition and internal energy modes.
	\item In the case of binning one group per internal energy mode, which is a particular case of (ii), we have the multi-temperature (MT) models~\cite{Park_Book_1990}.
\end{enumerate}
\par
\begin{figure}[!htb]
	\centering
	\includegraphics[width=0.4\linewidth]{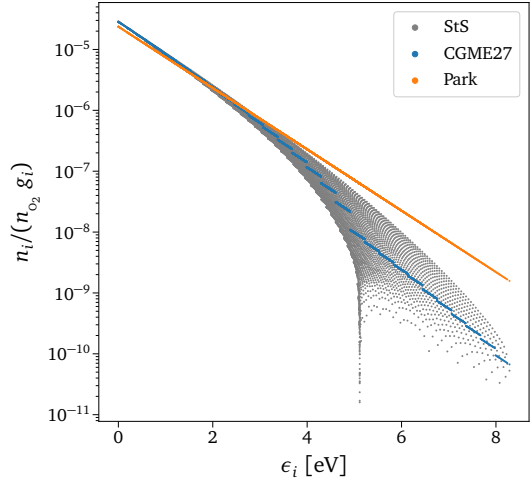}
	\caption{\textit{Normalized quasi-steady state (QSS) rovibrational states distribution for different models.} The level of physical accuracy can vary significantly depending on the choice of the thermochemical model. The orange dots are determined by low-fidelity Park's two-temperature model~\cite{Park_JTHT_2001}, \textit{e.g.}, a particular MT model, the blue ones by the 27-groups CG grouping strategy, and the grey ones by the high-fidelity StS modeling. Initial conditions used for the 0-$\mathrm{D}$ simulation: $P_{_0}=3\,000\text{ Pa}$, $X_{\text{O}_0}=0.2$, $T_{int_0}=1000\text{ K}$, $T=10\,000\text{ K}$.}
	\label{fig:norm_lev_dist_comparison}
\end{figure}

\FigRef{fig:norm_lev_dist_comparison} compares the levels of physical accuracy and resolution among the three models mentioned above for O$_2-$O kinetics, the only system considered in this work. A substantial loss of physical information can be noticed moving from the internal energy states distribution obtained with the StS model to the one defined by Park's two-temperature model~\cite{Park_Book_1990}, which is a particular case of the MT models, where all the states are collapsed along a straight line. Differently, the CGME approach better captures the StS distribution by modeling the dynamics of multiple clusters of states (27 in \figRef{fig:norm_lev_dist_comparison}, namely the CGME27 model). In this work, only the coarse-grained master equations approach will be employed to construct our surrogate, which is tested in both 0-$\mathrm{D}$ and 1-$\mathrm{D}$ scenarios.

\subsubsection{Coarse-grained modeling}\label{sssec:physics.CG}
The numerical solution of the master equations, whereby the dynamics of each internal energy state is captured via the direct solution of the corresponding mass conservation equations, is often impractically expensive. Moreover, it is usually not required since the internal energy distribution is generally a composition of partial equilibria rather than a complete non-equilibrium state~\cite{Sahai_JCP_2017}. The concept of local or partial equilibrium suggests the application of the \textit{principle of maximum entropy} to reconstruct the unresolved scales of physics~\cite{Panesi_JTHT_2011,Munafo_PF_2015,Liu_JCP_2015}.
The construction of a coarse-grained model is accomplished by adopting a two-step procedure which goes as follows~\cite{Munafo_AIAA_2020}:
\begin{enumerate}[i.]
	\item Group energy states into $\mathcal{N_G}$ macroscopic bins according to a specific strategy;
	\item Prescribe a bin-wise distribution function to represent the population within each group together with a series of moment constraints.
\end{enumerate}
This work employs a log-linear form for the bin-wise distribution function, which results in a thermalized local Boltzmann distribution within individual bins, defined as follows
\begin{equation}\label{eq:boltzmann}
	\mathcal{F}_P^i\left(\epsilon_i\right):\quad
	\log\left(\frac{g_i}{n_i}\right)=\alpha_P+\beta_P\epsilon_i\hspace{2mm},
\end{equation}
where the bin-specific coefficients $\alpha_P$ and $\beta_P$ are expressed in function of the macroscopic group constraints (\textit{i.e.}, number density, energy, \textit{etc.}). The total population and energies of the different bins are the set of unknowns for the reduced-order system. The governing equations for these macroscopic constraints can be derived by taking successive moments of the StS master equations, using $\left(\epsilon_i\right)^m$ for $m=0,1,\dots$ as weights (see Supplementary Section S.1.2 for more details).
\par
While more accurate strategies have been developed during the past few years~\cite{Sahai_JCP_2017,Venturi_JCP_2020}, the model-reduction approach employed in this work is the rovibrational energy-based grouping technique~\cite{Macdonald_I_JCP_2018,Macdonald_II_JCP_2018}, which lumps together energy states with similar internal energy regardless of their rotational and vibrational quantum numbers.

\subsubsection{Zero-dimensional chemical reactor}\label{sssec:physics.framework}
We wish to investigate the behavior of oxygen molecules in their electronic ground state undergoing dissociation when subjected to sudden heating in an ideal chemical reactor. We make the following assumptions:
\begin{enumerate}[i.]
	\item The 0-$\mathrm{D}$ reactor is plunged into a thermal bath maintained at constant temperature $T$.
	\item The translational energy mode of the atoms and molecules is assumed to follow a Maxwell-Boltzmann distribution at the temperature $T$ of the thermal bath.
	\item At the beginning of the numerical experiment, the population of the rovibrational energy levels is assumed to follow a Boltzmann distribution at the internal temperature $T_{int_0}$.
	\item The volume of the chemical reactor is kept constant during the experiment, and the thermodynamic system is closed (no mass exchange with the surrounding environment).
	\item Only $\alpha_P$ in \eqRef{eq:boltzmann} is modeled for each bin $P$, while $\beta_P=1/\left(k_BT_P\right)$ is kept constant during the 0-$\mathrm{D}$ simulation, with $k_B$ being the Boltzmann’s constant and $T_P=T$.
\end{enumerate}
Therefore, \eqRef{eq:gov_eq} reduces to
\begin{equation}\label{eq:gov_eq_0D}
	\begin{aligned}
		\frac{d\rho_i}{dt} &= \Omega^0_i \left(\rho_i,T\right) \\
		\rho_i\left(t=0\right) &= f_i
	\end{aligned}\hspace{2mm},
\end{equation}
where $f_i$ refers to the corresponding Maxwell-Boltzmann equilibrium value of specie $i$ at temperature $T_{int_0}$.
\par
Since the goal is to learn the integral solution operator of the rovibrational CG master equations to be able to deliver accurate predictions in multidimensional CFD simulations characterized by a wide range of physical scenarios, we aim to generalize over the space of initial conditions (ICs) and time domain. The ICs are generated by defining the initial pressure $P_{_0}$, the initial molar fraction of atomic oxygen $X_{\text{O}_0}$, the translational temperature $T$, and the initial internal temperature $T_{int_0}$ for which a Boltzmann distribution is prescribed for the O$_{2}$ bins. In this work, the domain in which the initial conditions have been sampled is defined in \tabRef{table:IC_space} as minimum-maximum pair values. For all the possible sampling scenarios, $T$ is greater than $T_{int_0}$, which implies that thermal excitation and dissociation processes are the dominant phenomena occurring in the reactor.
\par
\begin{table}[!htb]
	\centering
	\begin{tabular}{c|cccc}
		\toprule
		& $P_{_0}$ [Pa] & $X_{\text{O}_0}$ & $T_{int_0}$ [K] & $T$ [K] \\
		\midrule
		Min & 1\,000 & 0 & 1\,000 & 8\,000 \\
		Max & 10\,000 & 0.95 & 8\,000 & 15\,000 \\
		\bottomrule
	\end{tabular}
	\caption{\textit{Space of initial conditions.} Minimum-maximum pair values for each IC variable.}
	\label{table:IC_space}
\end{table}
Regarding the time domain, we train the model over an interval of [0,10$^{-2}$] s, covering most excitation and dissociation processes for the non-equilibrium problem under investigation.

\blue{
\subsubsection{One-dimensional numerical experiment}\label{sssec:shock}
Following the approach used by Zanardi \textit{et al.}~\cite{Zanardi_SciTech_2023}, a one-dimensional shock case scenario is employed to test the ML-based framework proposed in this work. The governing equations for the dynamics of inviscid, one-dimensional gas flows are given by the Euler equations:
\begin{equation}\label{eq:euler}
	\frac{\partial\mathbf{U}}{\partial t}
	+\frac{\partial\mathbf{F}}{\partial x}=\mathbf{S}\hspace{2mm},
\end{equation}
where $t$ represents time and $x$ represents space. It is worth noting that \eqRef{eq:gov_eq} is the Lagrangian version of \eqRef{eq:euler}, including an additional diffusive term. The vectors $\mathbf{U}$, $\mathbf{F}$, and $\mathbf{S}$ represent the conservative variables, inviscid fluxes, and source terms, respectively. They are defined as follows:
\begin{align}
	\mathbf{U}&=
	[\hspace{1mm}\rho_i\hspace{3mm}\rho u \hspace{3mm}\rho E\hspace{1mm}]^{\mathrm{T}}\hspace{2mm},\label{eq:euler.vec.a}\\
	\mathbf{F}&=
	[\hspace{1mm}\rho_iu \hspace{3mm}p +\rho u^2\hspace{3mm}\rho u H\hspace{1mm}]^{\mathrm{T}}\hspace{2mm},\label{eq:euler.vec.b}\\
	\mathbf{S}&=
	[\hspace{1mm}\Omega^0_i\hspace{3mm}0\hspace{3mm}0\hspace{1mm}]^{\mathrm{T}}\hspace{2mm},\label{eq:euler.vec.c}
\end{align}
where the total energy and enthalpy per unit-mass are $E = e + u^2/2$ and $H = E + p/\rho$, respectively. The thermodynamics of the system is explained in detail in the Supplementary Section S.1.1, and the variables $e$, $u$, $p$, and $\rho$ have their usual meanings in the context of gas dynamics. The source term $\Omega^0_i$ represents the mass production term, which is the same one as defined in \eqRef{eq:gov_eq_0D} and described in detail in the Supplementary Section S.1.2.
\par
The flow governing equations \eqref{eq:euler} are discretized in space using the finite volume method, with inviscid fluxes evaluated using van Leer's flux vector splitting in conjunction with the second-order upwind-biased MUSCL reconstruction procedure~\cite{VanLeer_JComP_1979, Hirsch_Book_2007}. The time integration method is based on the operator-splitting technique proposed by Strang~\cite{Strang_SIAM_1968}. This method integrates the transport operator, $\bm{\mathcal{T}}\left(\mathbf{U}\right)=\partial\mathbf{F}/\partial x$, and the reaction operator, $\bm{\mathcal{R}}\left(\mathbf{U}\right)=\mathbf{S}$, sequentially in a symmetric fashion:
\begin{align}
	\partial_t\mathbf{U}^{(1)} &=\bm{\mathcal{T}}\left(\mathbf{U}^{(1)}\right)\hspace{2mm}, & &
	\hspace{-5cm}\mathbf{U}^{(1)}\left(t_n\right)=\mathbf{U}_n \label{eq:split.01}\\
	\partial_t\mathbf{U}^{(2)} &=\bm{\mathcal{R}}\left(\mathbf{U}^{(2)}\right)\hspace{2mm}, & &
	\hspace{-5cm}\mathbf{U}^{(2)}\left(t_n\right)=\mathbf{U}^{(1)}\left(t_n+\Delta t/2\right) \label{eq:split.02} \\
	\partial_t\mathbf{U}^{(3)} &=\bm{\mathcal{T}}\left(\mathbf{U}^{(3)}\right)\hspace{2mm}, & &
	\hspace{-5cm}\mathbf{U}^{(3)}\left(t_n+\Delta t/ 2\right)=\mathbf{U}^{(2)}\left(t_n+\Delta t\right) \label{eq:split.03} \\
	\mathbf{U}_{n+1} &=\mathbf{U}^{(3)}\left(t_n+\Delta t\right)\hspace{2mm}, & & \label{eq:split.04}
\end{align}
where $\Delta t$ is the time step. The splitting formulation is second-order accurate, strongly stable, and symplectic for non-linear equations. Its convergence and stability properties have been extensively studied for reacting flow simulations~\cite{Knio_JComP_1999,Singer_CTM_2006,Ren_JComP_2014,Wu_CPC_2019}. The use of an operator-splitting approach facilitates the straightforward insertion of the constructed neural operator into the framework described by equations \eqref{eq:split.01} to \eqref{eq:split.04}. Instead of using an implicit scheme to integrate the stiff reaction step described by \eqRef{eq:split.02}, a simple evaluation of the trained surrogate is performed to evolve the solution in time. The surrogate takes the solution from the first flux integration step as input and provides the evolved gas state resulting from the reaction operator to the last step of the splitting scheme.

\paragraph{Test case configuration}
The main configuration details of the one-dimensional shock case scenario used herein are given below.
\begin{itemize}[-]
	\item \textit{Initial and boundary conditions}\\
	\TabRef{table:IC_1D} presents the piece-wise initial conditions. On the left side, freestream conditions corresponding to a hot gas at $T=3000$ K and $u=3000$ m/s are imposed. This choice is made because, at this temperature, the equilibrium state of the gas results in a reasonable amount of dissociated oxygen. It is important to note that this condition is not a requirement of the method itself but rather a consequence of only modeling the O$_2+$O kinetics without considering the O$_2+$O$_2$ system, where molecular oxygen alone is sufficient to activate the thermochemical processes. On the right side, the initial solution is set equal to the post-shock equilibrium state. A supersonic inflow boundary condition (BC) is imposed on the left side, where all characteristics are incoming, by prescribing all flow variables. A subsonic outflow BC is imposed on the right side with a specified pressure value.
	\begin{table}[!htb]
		\centering
		\begin{tabular}{c|c|cccc}
			\toprule
			& $x$ [m] & $u$ [m/s] & $p$ [Pa] & $X_{\text{O}}$ & $T$ [K] \\
			\midrule
			Freestream & $\le0.04$ & 3\,000 & 2\,940  & 0.12 & 3\,000 \\
			Post-shock (LTE) & $>0.04$ & 492 & 29\,598 & 0.39 & 4\,245 \\
			\bottomrule
		\end{tabular}
		\vskip 2mm
		\caption{Initial conditions for one-dimensional shock case scenario. }
		\label{table:IC_1D}
	\end{table}
	\item \textit{Time and space grid}\\
	The one-dimensional domain length is set to $L=0.1$ m, and the spatial discretization uses a space step of $\Delta x=4\times10^{-4}$ m, resulting in a total of 250 cells. The integration is performed until the shock profile is fully developed, using a total of 500 iterations with a constant time step of $\Delta t=1.33\times10^{-7}$ s determined by the freestream velocity while maintaining a maximum CFL number of 1 to ensure numerical stability.
\end{itemize}
\par
To ensure that the left and right equilibrium conditions are fully guaranteed and avoid any error accumulation due to even minor discrepancies in the surrogate's predictions, the inference is performed only for those cells experiencing non-local thermodynamic equilibrium (NLTE) effects, meaning for gas thermochemical states different from the ones shown in \tabRef{table:IC_1D}. However, to fairly compare the numerical integrator's and the surrogate's performance, the inference is performed for the whole 1-$\mathrm{D}$ domain, and the predictions for those cells in the same conditions as in \tabRef{table:IC_1D} are simply disregarded.
\par
To ensure physical consistency, the surrogate must learn the integral solution of the zero-dimensional formulation of \eqRef{eq:euler}, specifically \eqRef{eq:split.02}, which describes an adiabatic thermodynamic system without energy or mass exchange. Consequently, the isothermal assumption made in the 0-$\mathrm{D}$ analysis does not apply to this particular test. To accurately represent the adiabatic case, an additional DeepONet is required on top of the surrogate described in the next section. This additional DeepONet is employed to model the translational temperature $T$, enabling a more comprehensive and accurate representation of the complex thermochemical dynamics in the 1-$\mathrm{D}$ domain. Therefore, a distinct surrogate is constructed specifically for this simulation, with detailed information on data generation and network construction provided in the Supplementary Section S.3.1.
}

\subsection{Neural operators}\label{ssec:ml}
\subsubsection{DeepONet}\label{sssec:ml.opnn}
Building upon the original formulation of the DeepONet by Lu \textit{et al.}~\cite{Lu_NMI_2021}, whereby the solution operator $G$ maps an input function $\bm{u}$ and the continuous coordinates $\bm{y}$ of $G(\bm{u})$ to a real scalar value, this work extends the DeepONet framework to accommodate the high-dimensional nature of the master equations, thus obtaining an output vector $G(\bm{u})(\bm{y})\in\mathbb{R}^{D}$, where $D$ is the number of the output variables~\cite{Zanardi_AIAA_2022,Lu_CMAME_2022}. As illustrated in figure S1 in the Supplementary Information, the DeepONet architecture is characterized by two different deep neural networks: the ``branch net'' and the ``trunk net''. The modified version is characterized by multiple branches, one for each output variable, which takes $\bm{u}$ as input and returns a feature embedding $\bm{\alpha}\in\mathbb{R}^p$ as output. Instead, the trunk net takes the continuous coordinates $\bm{y}$ as inputs and outputs another feature embedding $\bm{\phi}\in\mathbb{R}^p$. This block is shared between different branches~\cite{Lu_CMAME_2022, Venturi_CMAME_2023}, gaining computational efficiency. In the framework of operator learning for ODEs, $\bm{u}$ represents the space of initial conditions, whereas $\bm{y}$ is the time variable. To obtain a continuous and differentiable representation of the output functions of the DeepONet, the outputs of each branch and the trunk networks are merged via dot product as follows:
\begin{equation}\label{eq:dot_prod}
	\widehat{G}^{(i)}(\bm{u})(\bm{y})=\sum_{k=1}^{p} \alpha^{(i)}_{k}(\bm{u}) \phi_{k}(\bm{y})
	\qquad\text{for}\hspace{2mm}i=1,\dots,D\hspace{2mm}.
\end{equation}
One can notice that \eqRef{eq:dot_prod} reminds the proper orthogonal decomposition (POD) formulation~\cite{Berkooz_ARFM_1993}, as highlighted by Lu \textit{et al.}~\cite{Lu_CMAME_2022}, and more generally \eqRef{eq:dot_prod} can be related to the singular value decomposition (SVD) factorization, as explained by Venturi and Casey~\cite{Venturi_CMAME_2023}. From this perspective, the trunk net learns the $p$ most important modes of the dynamical system, $\bm{\phi}$, while the branch net learns the coefficients $\bm{\alpha}$ of the expansion. Under this perspective, the shared-trunk version of the DeepONet works reasonably well only when the dynamics of the modeled variables are similar to each other such that they can share the same basis $\bm{\phi}$~\cite{Venturi_CMAME_2023}.

\subsubsection{Multi-scale hierarchical coarse-grained model}\label{sssec:ml.architecture}
Similar to what is done in adaptive mesh refinement (AMR) techniques used in CFD, the accuracy of the CG model can be improved by increasing the number of groups but at a higher computational cost. The improvement in accuracy is explained by the larger range of scales (or kinetic processes) that can be resolved. Indeed, taking as an example the rovibrational energy-based grouping strategy employed in this work, if we recursively split the energy space of the internal states by following a cascade in the groups, all the micro-groups inside the corresponding macro-group quickly reach the same equilibrium value, showing a fast dynamical behavior. Consistently, we leveraged the multi-scale nature of the physical problem to construct a physics-inspired ML-based surrogate (see Supplementary Section S.2.2 for all the details) by sequentially learning the different timescales of the thermochemical phenomena occurring inside a 0-$\mathrm{D}$ reactor.
\begin{itemize}[-]
	\item \textit{Timescale 1}\\
	Chemical dissociation of O$_2$ molecules (irrespective of their internal excitation) and creation of O atoms are the slowest processes that can be learned. As shown in \figRef{fig:hierarchical.arch}(a), the outputs of the DeepONet employed for this first timescale, denominated as CG-DeepONet$^{(1,1)}$ (\textit{i.e.}, the surrogate's component in charge of predicting the group number one in the scale number one), are simply the mass fractions of O and O$_2$. So, we are assuming that all the internal states can be clustered in one unique group, but we do not solve for the rovibrational-translation energy transfer phenomena. As concerns the physical input of the model, $\bm{u}$ represents the initial conditions of the reactor, which is characterized by translational temperature, $T$, reactor density, $\rho$, and initial mass fraction of O$_2$, while the independent variable, $\bm{y}$, of the operator $G\left(\bm{u}\right)$ is the time, $t$:
	\begin{equation}\label{eq:TS1_IO}
		\begin{aligned}
			\bm{u}&=\left[T,\,\rho,\,Y_{{\text{O}_2}_{0}}\right]&&\quad\in\mathbb{R}^{3}\\
			\bm{y}&=t&&\quad\in\mathbb{R}^1\hspace{2mm}.\\
			\widehat{G}(\bm{u})(\bm{y})&=\left[\widehat{Y}_{\text{O}}\vert_{\bm{u}}(t),\;
			\widehat{Y}_{\text{O}_2}^{(1,1,1)}\vert_{\bm{u}}(t)\right]&&\quad\in\mathbb{R}^2
		\end{aligned}
	\end{equation}
	In \eqref{eq:TS1_IO} and \figRef{fig:hierarchical.arch}(a), a series of two or three superscripts have been used, where the first one corresponds to the \textit{timescale} investigated, the second the DeepONet \textit{index}, and the last one the O$_2$ \textit{group}. They will help to identify the different variables and DeepONets used for each timescale. The \textit{Softmax} function in \figRef{fig:hierarchical.arch}(a) is applied to the dot product outputs after these being linearly transformed. It guarantees the mass fractions to be positive values and the mass to be conserved.
	\begin{figure}[!t]
		\centering
		\includegraphics[width=0.8\linewidth]{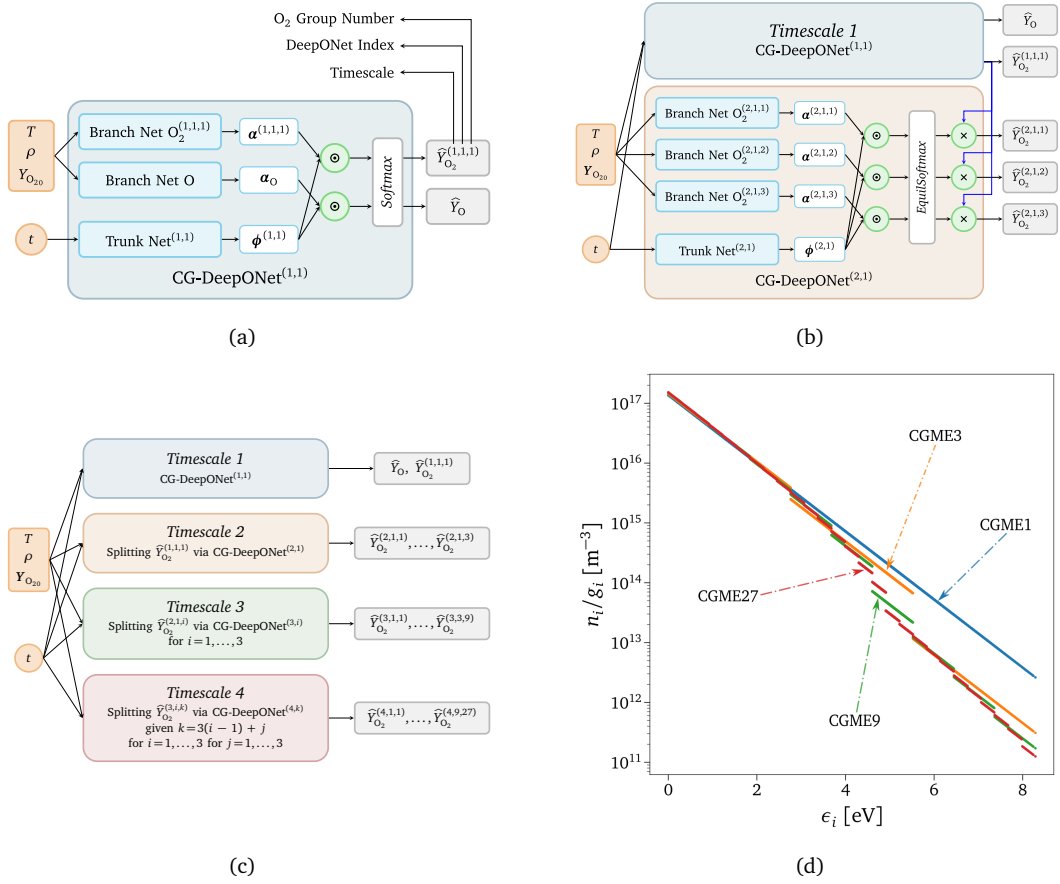}
		\caption{\textit{Hierarchical surrogate for multi-scale coarse-grained dynamics.} (a) Schematics of the network architecture for 1-group CG master equations (CGME1). (b) Schematics of the network architecture for 3-groups CG master equations with the additional \textit{EquilSoftmax} layer. (c) Schematics of the complete hierarchical network architecture. (d) O$_2$ rovibrational distribution for 1-group (CGME1), 3-groups (CGME3), 9-groups (CGME9), and 27-groups (CGME27) coarse-grained grouping.}
		\label{fig:hierarchical.arch}
	\end{figure}
	\item \textit{Timescale 1-2}\\
	In the following timescale, we start modeling the energy exchange processes for O$_2$. To do so, the internal states are clustered into three groups (CGME3) which is equivalent to uniformly splitting the energy space covered by the unique group from the previous timescale (CGME1) into three parts, as shown in \figRef{fig:hierarchical.arch}(d). To learn the dynamics of this new system, the information learned from the previous timescale is leveraged by adopting transfer learning for the calibrated weights of CG-DeepONet$^{\left(1,1\right)}$. The new DeepONet is designed to learn the 3-group normalized distribution. The mass fractions of the three bins are then obtained by multiplying the modeled distribution by the total mass fractions of O$_2$ predicted by CG-DeepONet$^{\left(1,1\right)}$, as shown in \figRef{fig:hierarchical.arch}(b), ensuring the conservation of mass across the two scales. In terms of architecture, two are the difference between \textit{Timescale 1} and \textit{Timescale 2}. The first is related to the inputs, $\bm{u}$, of the branch net, which considers the initial mass fractions of all the three groups, $\boldsymbol{Y}_{{\text{O}_2}_{0}}$. Since \textit{Timescale 1} takes as an input the total mass fraction of O$_2$ as described in \eqref{eq:TS1_IO}, the three values are summed to get the correct input for CG-DeepONet$^{\left(1,1\right)}$. The second aspect concerns the replacement of the \textit{Softmax} layer with the \textit{EquilSoftmax} one. The latter can be considered as an extension of the former, and it has the following formulation:
	\begin{equation}\label{eq:EquilSoftmax}
		\frac{\widehat{Y}_{\text{O}_2}^{(2,1,i)}}{\widehat{Y}_{\text{O}_2}^{(1,1,1)}}
		=\textit{EquilSoftmax}\left(\bm{x}\right)_i=\frac{\exp\left(x^{(2,1,i)}\right)Q_i(T)}
		{\sum\limits_{i}\exp\left(x^{(2,1,i)}\right)Q_i(T)}
		\qquad\text{for}\hspace{2mm}i=1,2,3\hspace{2mm},
	\end{equation}
	where $Q_i(T)$ is the internal partition function of group $i$. Therefore, if $x^{(2,1,i)}=0$ $\forall$ $i$, all the groups are in equilibrium at the translational temperature $T$. In the case of isothermal reactors, $T$ is provided as one of the inputs $\bm{u}$. Conversely, for adiabatic systems like the 1-$\mathrm{D}$ test case scenario considered in this work, $T$ is predicted by a separate DeepONet. This additional transformation layer, referred to as the Boltzmann layer in the introductory section, enforces local equilibrium distributions between states in the same group by construction. Moreover, it positively impacts the regularization of the network by providing a physically consistent prior distribution to anchor the network parameters, specifically a zero-valued distribution, which can be effectively regulated using $L^2$ regularization. This ensures that the surrogate predictions remain closely aligned with the known reference equilibrium state, preventing excessive divergence and enhancing the robustness of the surrogate. It is worth highlighting that during the joint training process, all the parameters of CG-DeepONet$^{\left(1,1\right)}$ are re-trained together with the ones of CG-DeepONet$^{\left(2,1\right)}$, rather than being kept frozen. This is performed by employing fine-tuning transfer learning with \textit{L$^1$-SP} and \textit{L$^2$-SP} regularization as described in \refCite{Li_PR_2020}.
	\item \textit{Faster Timescales}\\
	It is possible to increase the accuracy of the CG model by further splitting the energy space into a higher number of clusters. Therefore, by sequentially repeating the same procedure that has been done for augmenting the model from \textit{Timescale 1} to \textit{Timescale 2}, we can construct a surrogate that can predict the dynamics of high-resolution CG models. In our case, we further split each bin into three more bins, obtaining first a 3-group CG modeling for \textit{Timescale 2}, then a 9-group CG modeling for \textit{Timescale 3}, and finally a 27-group  CG modeling for \textit{Timescale 4}. We treat each group's triplet with a single DeepONet, and we apply the \textit{EquilSoftmax} layer at the output of each entire timescale block. As explained in the previous paragraph, the predicted mass fraction of each macro-group multiplies the distribution of the corresponding three micro-groups, obtaining a hierarchical surrogate for multi-scale coarse-grained dynamics, as shown in \figRef{fig:hierarchical.arch}(c).
\end{itemize}

\subsubsection{Training strategy}\label{sssec:train}
Physics-informed neural networks (PINNs)~\cite{Raissi_JComP_2019} can integrate data and physical governing laws by adding PDE residuals to the loss function of neural networks by relying on automatic differentiation. This capability can also be incorporated into the DeepONet framework (physics-informed DeepONet or PI-DeepONet)~\cite{Wang_SA_2021,WangPerdikaris_arxiv_2021}. Specifically, the following composite loss function is minimized to train the network parameters, $\boldsymbol{\theta}$:
\begin{equation}\label{eq:loss}
	\mathcal{L}(\bm{\theta})=\lambda_{d}\mathcal{L}_{d}(\bm{\theta})+\lambda_{r}\mathcal{L}_{r}(\bm{\theta})
	+\lambda_{ic}\mathcal{L}_{ic}(\bm{\theta})+\Lambda(\bm{\theta})\hspace{2mm},
\end{equation}
where $\mathcal{L}_{d}(\bm{\theta})$ is computed based on the discrepancy between predicted and given data points, $\mathcal{L}_{r}(\bm{\theta})$ is the residual loss, $\mathcal{L}_{ic}(\bm{\theta})$ is the loss over the initial conditions of the 0-D reactor, and $\Lambda(\bm{\theta})$ contains the $L^1$ and $L^2$ regularization loss. These terms can be expressed as follows:
\begin{align}
	\mathcal{L}_{d}(\bm{\theta})
	&=\frac{1}{N_{d}}\sum_{i=1}^{N_{d}}\ell\left(\bm{Y}\left(\bm{u}^i_d,t^i_d\right),\bm{\widehat{Y}}\left(\bm{u}^i_d,t^i_d\right)\right)\hspace{2mm},\\
	\mathcal{L}_{r}(\bm{\theta})
	&=\frac{1}{N_{r}}\sum_{i=1}^{N_{r}}r
	\left(t^i_r,\rho^i_r,\bm{\widehat{Y}}\left(\bm{u}^i_r,t^i_r\right),T^i_r\right)\hspace{2mm},\\
	\mathcal{L}_{ic}(\bm{\theta})
	&=\frac{1}{N_{ic}}\sum_{i=1}^{N_{ic}}\ell\left(\bm{Y}\left(\bm{u}^i_{ic},0\right),
	\bm{\widehat{Y}}\left(\bm{u}^i_{ic},0\right)\right)\hspace{2mm},
\end{align}
where $N_{d}$, $N_{r}$, and $N_{ic}$ denote the batch sizes of the training data. $\bm{Y}$ are the exact mass fraction values from direct numerical simulation of the CG master equations (CGME), whereas $\bm{\widehat{Y}}$ are the predicted ones from the surrogate. The parameters $\lambda_d$, $\lambda_r$, and $\lambda_{ic}$ correspond to weight coefficients in the loss function that can effectively assign a different learning rate to each loss term. In this study, the error function $\ell$ is expressed as follows:
\begin{equation}
	\ell\left(\bm{Y},\bm{\widehat{Y}}\right)
	=\frac{1}{|\bm{\widehat{Y}}|}\left\Vert\frac{\ln\left(\bm{Y}\right)
		-\ln\left(\bm{\widehat{Y}}\right)}{\ln\left(\bm{Y}\right)}\right\Vert_1\hspace{2mm},
\end{equation}
while the residual $r\in\mathbb{R}$ is
\begin{equation}\label{eq:res}
	r\left(t,\rho,\bm{\widehat{Y}},T\right)
	=\frac{1}{|\bm{\widehat{Y}}|}\left\Vert\rho\frac{d\bm{\widehat{Y}}}{dt}
	-\boldsymbol{\Omega^0}\left(\rho,\bm{\widehat{Y}},T\right)\right\Vert_1\hspace{2mm},
\end{equation}
with $\bm{\Omega^0}$ being the right hand side of \eqRef{eq:gov_eq_0D}.
\par
\violet{
Given the hierarchical structure of the proposed surrogate, the parameters of the entire network are trained by adopting a multi-step procedure:
\begin{enumerate}[i.]
	\item \textit{Fully data-driven optimizations}\\
	In this first step, the surrogate is trained sequentially from the slowest to the fastest timescale with only anchor and ICs points $\left(\lambda_d=1,\,\lambda_r=0,\,\lambda_{ic}=1\right)$ obtained from the numerical solution of the coarse-grained master equations:
	\begin{compactenum}[a)]
		\item Training only \textit{Timescale 1} with data generated by solving CGME1;
		\item Training jointly \textit{Timescales 1-2} with data generated by solving CGME3;
		\item Training jointly \textit{Timescales 1-2-3} with data generated by solving CGME9;
		\item Training jointly \textit{Timescales 1-2-3-4} with data generated by solving CGME27.
	\end{compactenum}
	At each training step, the knowledge acquired from the previous iterations is preserved and used as a prior by employing fine-tuning transfer learning with \textit{L$^1$-SP} and \textit{L$^2$-SP} regularization as described in \refCite{Li_PR_2020}. For instance, in step (b), the calibrated weights for \textit{Timescale 1} from step (a) are kept and finely retrained with the newly initialized parameters of \textit{Timescale 2}.
	\item \textit{Hybrid physics-informed and data-driven optimization}\\
	The governing equations describing the CGME27 model are now enforced in the trained surrogate from step (i.d) using the hybrid loss formulation shown in \eqRef{eq:loss}. The weight coefficients $\lambda_i$ are automatically tuned using the learning rate annealing technique described in \refCite{Wang_arXiv_2020}. The tuning procedure involves balancing the gradients of different loss terms during back-propagation using $\lambda_i$ as a re-scaling factor of the learning rate corresponding to each loss term. This technique ensures that the model's parameters are updated in a balanced manner, giving equal importance to all the loss terms. The complete training history of the parameter values $\lambda_i$ can be found in the Supplementary Section S.2.2.3.
\end{enumerate}
The decision to incorporate the residual loss only in the final step is intended to accelerate the training of the entire surrogate. Data from numerical simulations serves as anchor points for frequent or commonly seen scenarios, while the residual of the governing laws ensures the model's ability to generalize to different, unseen physical conditions.
}

\subsection{Adaptive pruning technique}\label{ssec:ml.adapt_tech}
\begin{figure}[!t]
	\centering
	\includegraphics[width=0.8\linewidth]{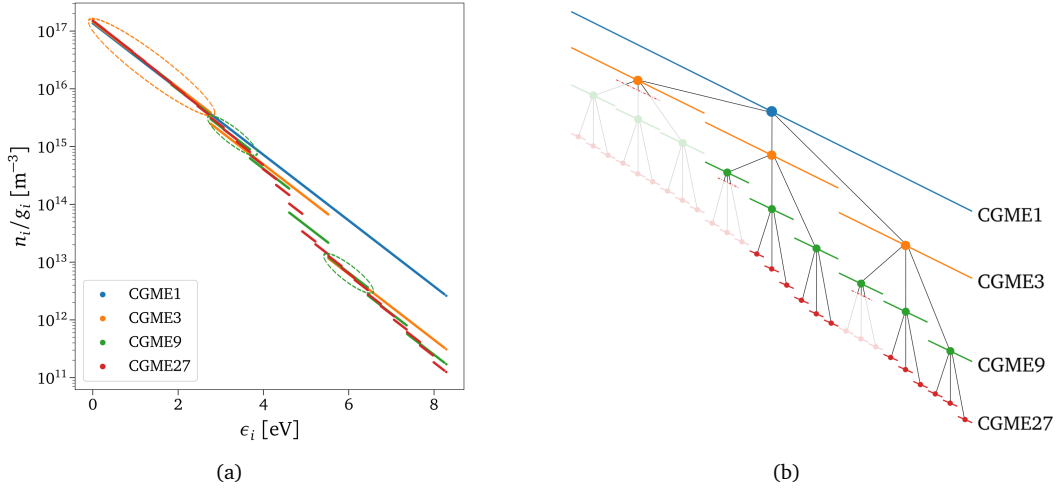}
	\caption{\textit{Example of the adaptive strategy.} (a) O$_2$ QSS rovibrational distribution for 1-group (CGME1), 3-groups (CGME3), 9-groups (CGME9), and 27-groups (CGME27) coarse-graining. The dashed ovals identify those CG high-resolution groups that can be accurately reconstructed from the low-resolution ones. (b) Exploded view of the groups' graph. The opaque dots represent the Boltzmann-reconstructed groups that correspond to the ovals in (a) and do not require evaluations of CG-DeepONets' high-resolution components.}
	\label{fig:adapt_example}
\end{figure}

Flow simulations are often characterized by regions of strong and weak non-equilibrium conditions of the gas. When the extent of non-equilibrium is large, the highest resolution is needed to resolve all the physical processes accurately. However, there are conditions for which the fine scales (or micro-groups) corresponding to the highest resolution CG model are in equilibrium with other neighboring groups or states. For these cases, adding resolution penalizes the computational efficiency rather than improving the model's accuracy. In fact, under these conditions, the population distribution can be approximated with a Boltzmann distribution, and the low-fidelity CG model can accurately resolve their dynamics. \FigRef{fig:adapt_example} illustrates the concept described above, where all the reconstructed low-lying energy states from different coarse-grained (CG) models are considered to be in equilibrium. As a result, it is sufficient to predict the values of the first group of the CGME3 model, without needing to resolve all the timescales.
These observations indicate the need to introduce a controller in the algorithm that accurately determines the resolution level needed to describe the dynamics of the system, without explicitly computing unnecessary fine scales. \blue{In the following, the design procedure for the additional controller-acting surrogate is firstly outlined, including the definition of the control variable and the network architecture. Subsequently, the adaptive inference technique is described, which involves the dynamic pruning of unnecessary nodes in the CG-DeepONets hierarchical architecture. This online pruning process enhances computational efficiency by selectively skipping the evaluation of specific nodes based on the local thermochemical state of the gas.
\begin{itemize}[-]
	\item \textit{Physically-relevant non-equilibrium control variable} \\
	First, defining a metric that can quantify the physical information lost due to the coarse-graining procedure is crucial. This work employs the Euclidean distance between the Boltzmann reconstructed states of the highest resolution CG model available (\textit{i.e.}, \textit{Timescale 4}) and the remaining low-fidelity ones. Since only the zeroth-order moment of the master equations is considered, the bin-specific coefficient $\alpha$ in \eqRef{eq:boltzmann} is selected to construct our metric, which can be expressed as follows:
	\begin{equation}\label{eq:euclidean_metric}
		\delta^{\left(ts,\cdot,P\right)}=\frac{1}{\mathcal{N}_p}
		\sum_{p,\mathcal{I}^{\left(4,\cdot,p\right)}\subset\mathcal{I}^{\left(ts,\cdot,P\right)}}
		\left(\alpha^{\left(ts,\cdot,P\right)}-\alpha^{\left(4,\cdot,p\right)}\right)^2\hspace{2mm},
	\end{equation}
	where $ts$ and $P$ (or $p$) refer to the timescale and its specific group, respectively. \EqRef{eq:euclidean_metric} involves the computation of the difference between the offsets of the log-linear Boltzmann distribution functions described in \eqRef{eq:boltzmann}. The sum in \eqRef{eq:euclidean_metric} is performed over all the $\mathcal{N}_p$ micro-groups of \textit{Timescale 4} that belong to the macro-group $P$ of timescale $ts$. \FigRef{fig:adapt}(a) provides a visual intuition of \eqRef{eq:euclidean_metric} for the first CGME3-group, which consists of the sum of the drawn dashed black lines. We briefly mention that other options for constructing the metric could have relied on the Kullback-Leibler divergence computed between population or energy distributions at the different temporal scales.
	\item \textit{Controller-acting surrogate architecture} \\
	Given the defined metric, the design of the non-equilibrium controller-acting surrogate requires a specific architecture. To maintain consistency with the coarse-grained operator network described in \SecRef{sssec:ml.architecture}, we again leverage the multi-scale connotation of the physical problem by separately modeling the underpredicted non-equilibrium values for each CG low-fidelity model, as illustrated in \figRef{fig:adapt}(b). An exponential transformation is applied to the surrogate outputs, and a single DeepONet is used for each triplet of values, following a similar approach as used for the CG-DeepONets. More details can be found in the Supplementary Section S.2.3.
	\begin{figure}[!htb]
		\centering
		\includegraphics[width=0.8\linewidth]{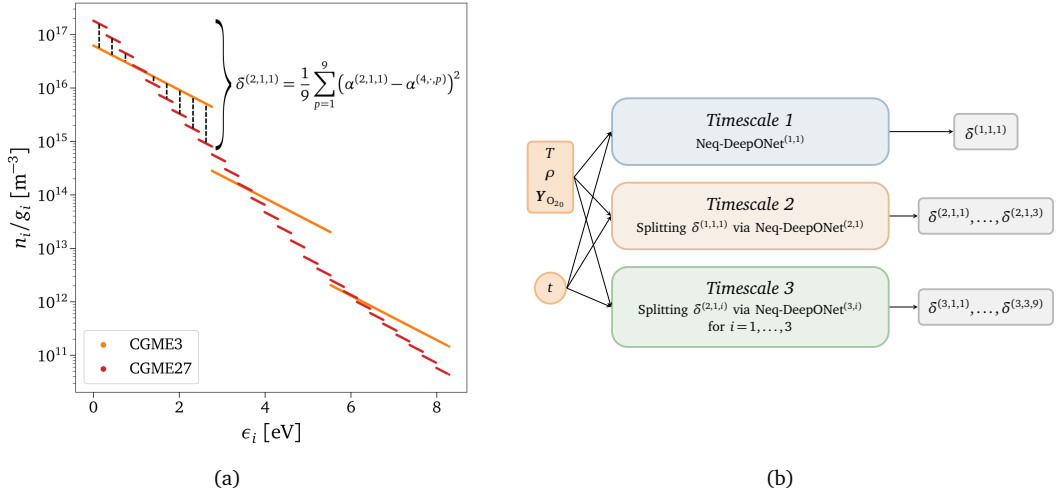}
		\caption{\textit{Adaptive inference design.} (a) Euclidean distance metric, $\delta$, used to quantify the physical information lost due to the equilibrium assumption imposed in a too-large subspace in the energy phase. $\alpha$ represents the zeroth-order term, \textit{i.e.}, the offset of the log-linear Boltzmann distribution function defined in \eqRef{eq:boltzmann}. (b) Schematics of the multi-scale network architecture of the controller-acting surrogate responsible for adapting the required coarse-grained model resolution based on the local flow conditions.}
		\label{fig:adapt}
	\end{figure}
	\item \textit{Physics-driven online pruning} \\
	The composition of coarse-grained deep operator networks (CG-DeepONets) and non-equilibrium controller-acting DeepONets (Neq-DeepONets) allows the development of a technique that, given IC and time instant, adaptively predicts the groups’ distribution with the highest accuracy and lowest computational cost possible. This technique can be summarized as a two-step procedure which goes as follows:
	\begin{enumerate}[i.]
		\item The first step involves querying the Neq-DeepONets to obtain the non-equilibrium control variable $\delta$ for each CG resolution level. This variable reflects the inaccuracy of the low-fidelity CG models in describing the non-equilibrium state of the gas at the upcoming time instant.
		\item The predicted $\delta$ is then compared with a user-chosen tolerance level, $\delta_{\text{tol}}$. If the predicted value is lower than the tolerance, the resolution level of the specific low-fidelity CG model is deemed sufficient to accurately represent the reactor dynamics. In such a case, the leaf nodes of the corresponding dependent tree in the CG-DeepONets model are temporarily pruned and not evaluated, as exemplified in \FigRef{fig:adapt_example}(b).
	\end{enumerate}
	At this point, we highlight the twofold advantage of CG-DeepONets’ hierarchical structure. In fact, other than simplifying the training stage, the presence of the controller boosts the inference phase, as the surrogate relies only on the CG-DeepONets’ components that are truly required to characterize the non-equilibrium distributions. The details of the adaptive algorithm are presented in the Supplementary Section S.2.3.
\end{itemize}
}

\section*{Results}\label{sec:results}
The framework discussed in the previous sections is used to construct a surrogate for an ideal chemical reactor. The first part of this section provides the details of the training and testing of the surrogate in isothermal 0-D scenarios, demonstrating its ability to learn the differential operator governing the physics of the reactor. The surrogate's predictions are then compared against the solutions obtained from the numerical integration of the governing equations. Observables such as time-resolved distributions and its moments, including densities and energies, are employed for evaluation. Furthermore, details regarding the adaptive technique and a preliminary analysis of computational savings are provided.
At the end of the section, the results of the one-dimensional numerical experiment are analyzed in terms of surrogate accuracy and performance.

\subsection{Inference}\label{ssec:res.inference}
\begin{figure}[!htb]
	\centering
	\includegraphics[width=0.45\linewidth]{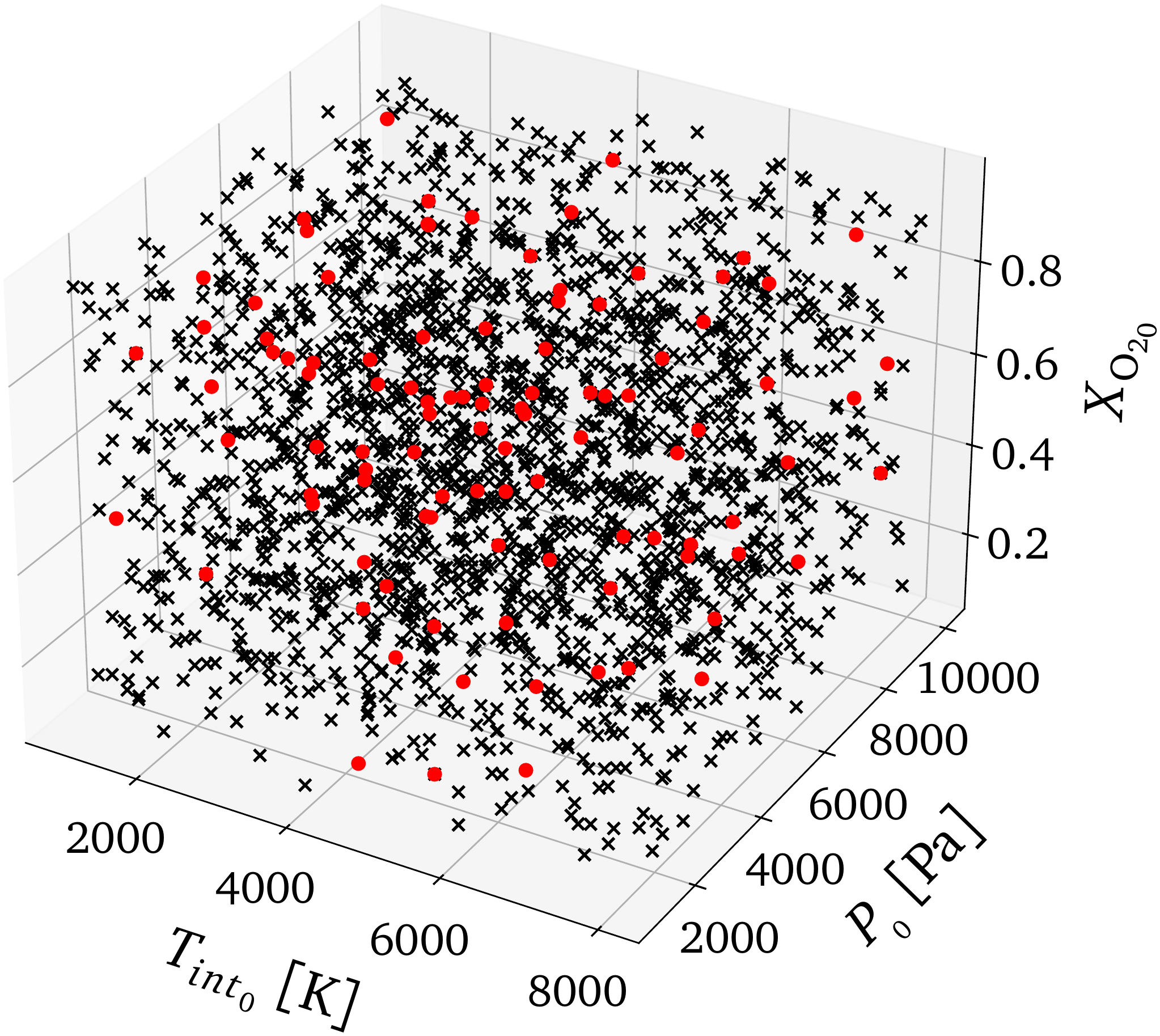}
	\caption{\textit{Space of initial conditions.} The black crosses represent the set of training points, while the red dots identify the testing data set. Note: the figure is missing the last fourth dimension in the space of initial conditions, \textit{i.e.}, the translational temperature of the reactor, $T$.}
	\label{fig:cloud}
\end{figure}

As explained in \SecRef{sssec:physics.framework}, different initial conditions have been uniformly sampled from \tabRef{table:IC_space} to train and test the proposed ML framework. \FigRef{fig:cloud} shows the broad ranges of the space of ICs for pressure, $P_{_0}$, molar fraction of atomic oxygen, $X_{\text{O}_0}$, and internal temperature, $T_{int_0}$. A fourth dimension should be considered since the translational temperature of the reactor, $T$, also varies. In \figRef{fig:cloud}, the red dots represent unseen test scenarios, whereas the black crosses represent the training points.
\par
\begin{figure}[!htb]
	\centering
	\includegraphics[width=0.8\linewidth]{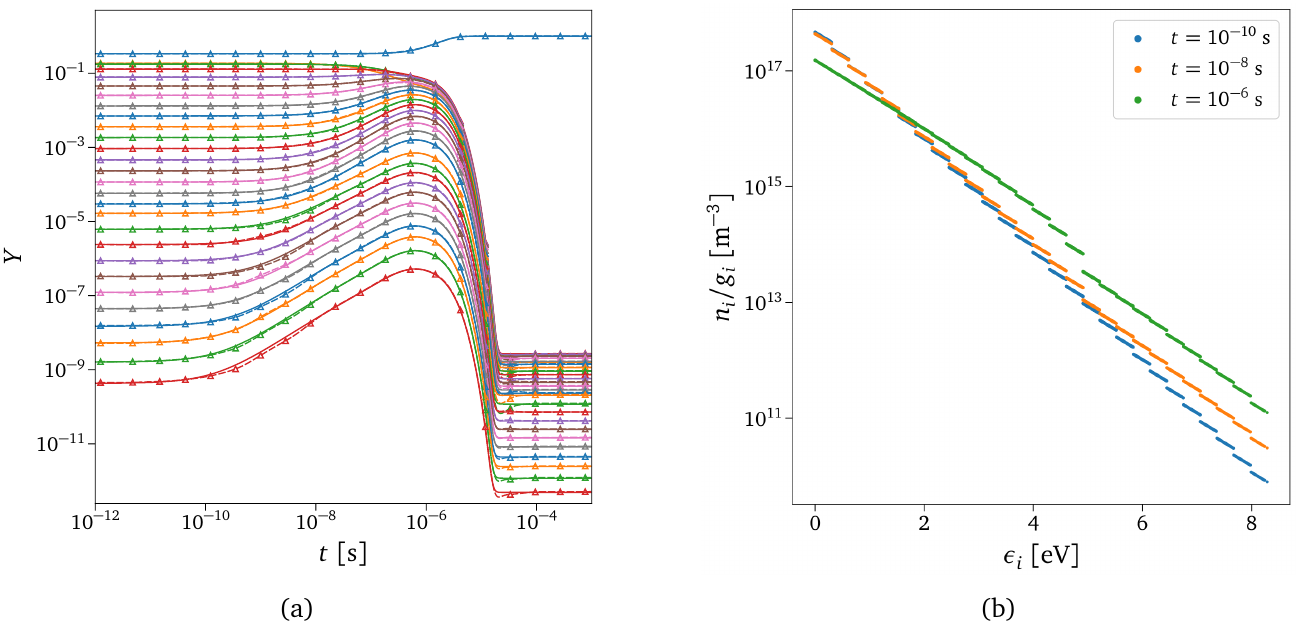}
	\caption{\textit{Inferred solution.} (a) Reference solution (solid line) versus prediction from the trained model (dashed line with markers). The isolated blue line represents the evolution of the atomic oxygen, while the remaining are the 27 groups of O$_2$ in ascending order of energy content per particle (top-down). Initial conditions used: $P_{_0}=3\,617.24\text{ Pa}$, $X_{\text{O}_0}=0.4996$, $T_{int_0}=4\,333.16\text{ K}$, $T=9\,742.13\text{ K}$. (b) Predicted O$_2$ rovibrational states distribution at three different time instants taken from the dynamics shown in (a).}
	\label{fig:inference}
\end{figure}

\FigRef{fig:inference}(a) compares the exact solution computed by the numerical integrator and the surrogate's predictions for one unseen scenario taken from the test data set in \figRef{fig:cloud}. The isolated blue line represents the evolution of the atomic oxygen taken from \textit{Timescale 1}. In contrast, the others describe the dynamics of the 27 rovibrational energy-based groups predicted by \textit{Timescale 4}. The inference has been performed by querying the CG-DeepONet based on the vector of time instants generated from the numerical integrator and the given initial conditions, defined by $\left\{\left[T,\,\rho,\,\bm{Y}_{{\text{O}_2}_{0}}\right],\,t_k\right\}^{M}_{k=1}$, with $M$ the number of evaluation points.
From \figRef{fig:inference}(a), it can be observed that the predicted and exact solutions show excellent agreement. This indicates that the trained model is capable of accurate predictions for different and unseen initial conditions (additional test cases are presented in the Supplementary Section S.2.2.2). Negligible discrepancies can be noticed in various regions of the dynamics of the heat bath, which can be improved by further refining the trained model. To the author’s best knowledge, this work provides the first application of PI-DeepONets to a dynamical system containing many such degrees of freedom. \red{The main reason for such good surrogation of the dynamics is that the hierarchical structure of the proposed deep learning framework embodies the multi-scale connotations of the problem, showing higher accuracy and robustness compared to a vanilla DeepONet architecture (details provided in the Supplementary Section S.2.1.1).} The micro-groups inside each macro-groups equilibrate faster between each other than with other ones outside it. For this reason, they show very similar behavior in their dynamics, which can be captured by the few modes discovered by the shared trunk. This aspect facilitates reaching high levels of accuracy with a relatively small number of network parameters. Indeed, the surrogate correctly predicts the dynamics of almost thirty species spanning a wide range of orders of magnitude (around 12) in mass fractions values. Additionally, to expand the initial conditions’ space even further by keeping such a high accuracy level and relatively small network architecture, one could consider constructing multiple surrogates. Each of these surrogates can be built with the same architecture but specialized for a local sub-domain in the space of the initial conditions.

\subsection{Accuracy}\label{ssec:res.accuracy}
The relative $L^2$-norm has been used as the error metric to evaluate the accuracy of the surrogate, consistently with \refCite{Wang_SA_2021}. In particular, the employed test error corresponds to the mean relative error of the surrogate's predictions for \textit{Timescale 4} over all the examples in the test data set:
\begin{equation}
	\varepsilon^{(j)}=\frac{1}{N} \sum_{i=1}^{N}
	\dfrac{\left\|\widehat{Y}_i^{(4,\cdot,j)}(t) - Y^{(4,\cdot,j)}_i(t)\right\|_2}
	{\left\|Y^{(4,\cdot,j)}_i(t)\right\|_2}
	\qquad\text{for }\hspace{1mm}j\in\{1,\dots,\mathcal{N_G}\}\hspace{2mm},
\end{equation}
where $\mathcal{N_G}=27$ represents the number of groups, $N=100$ denotes the number of testing cases, and $t$ represents a set of log-uniformly spaced points in the time domain. For this analysis, $1\,000$ points in time have been sampled from each testing scenario. The four highest errors of the inferred solution are presented in \tabRef{table:error}. Once again, the reported values confirm the excellent agreement between the numerically integrated master equations and the predicted solutions, with a maximum relative $L^2$-norm error of approximately 4.5\%.
\begin{table}[!htb]
	\centering
	\renewcommand{\arraystretch}{1.6}
	\begin{tabular}{c|c}
		\toprule
		Group & Rel. error [\%] \\
		\midrule
		$\widehat{Y}^{(4,2,6)}$ & $4.52\pm2.44$ \\
		$\widehat{Y}^{(4,3,9)}$ & $4.12\pm2.64$ \\
		$\widehat{Y}^{(4,9,27)}$ & $3.82\pm1.59$ \\
		$\widehat{Y}^{(4,6,21)}$ & $3.35\pm1.54$ \\
		\bottomrule
	\end{tabular}
	\caption{\textit{Test error.} The four highest mean relative $L^2$-norm testing errors (with standard deviations) of the trained model for \textit{Timescale 4}.}
	\label{table:error}
\end{table}

\subsection{Surrogate predictions \textit{vs.} numerically-integrated thermochemical models}\label{ssec:res.surrogate_vs_park}
\begin{figure}[!htb]
	\centering
	\includegraphics[width=0.8\linewidth]{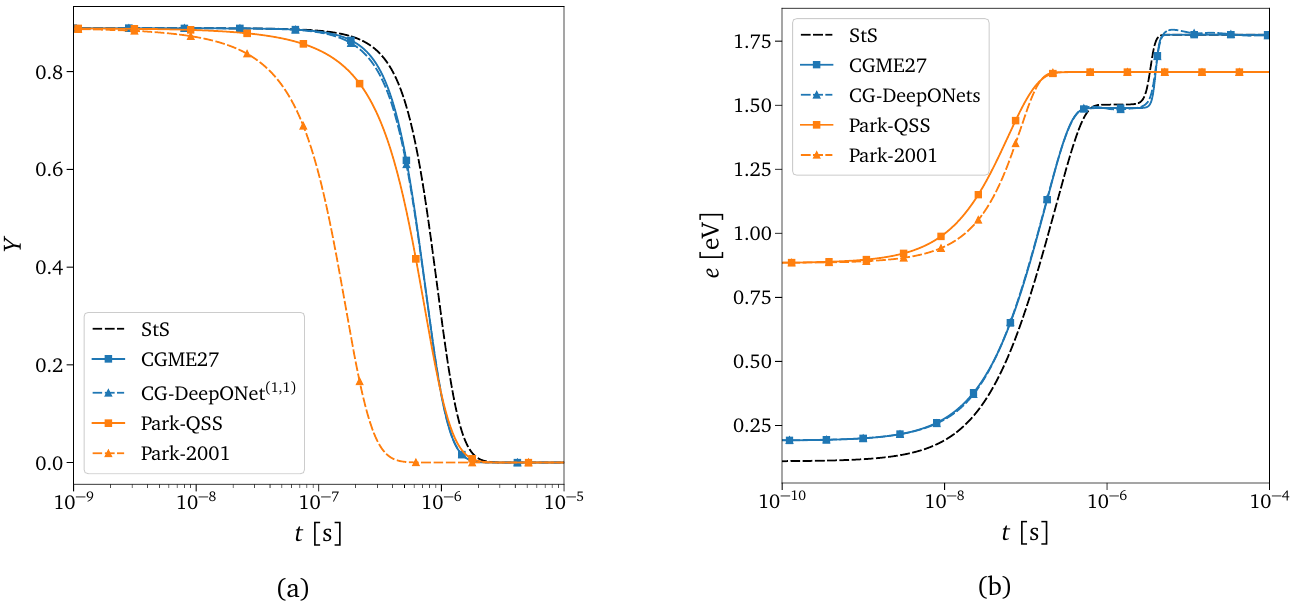}
	\caption{\blue{\textit{Surrogate predictions compared to numerically-integrated thermochemical models.}
	\tinybullet Black dashed line: state-to-state exact solution
	\tinybullet Blue line with markers: coarse-grained exact solution
	\tinybullet Blue dashed line with markers: coarse-grained surrogate predictions
	\tinybullet Orange line with markers: Park's two-temperature model with QSS approach~\cite{Panesi_JCP_2013}
	\tinybullet Orange dashed line with markers: Park's two-temperature model with kinetics from \refCite{Park_JTHT_2001}. Initial conditions used: $P_{_0}=3\,000\text{ Pa}$, $X_{\text{O}_0}=0.2$, $T_{int_0}=1000\text{ K}$, $T=10\,000\text{ K}$. (a) Evolution of total mass fraction of O$_2$. (b) Weighted rovibrational energy evolution per particle of O$_2$.}}
	\label{fig:park_vs_TS1}
\end{figure}
\blue{To demonstrate the level of physical accuracy of the coarse-grained surrogate discussed in this study, a comparison is made against the reference CG solution, the high-fidelity state-to-state solution, and the computationally cheaper two-temperature model of Park, which is a specific case of the multi-temperature models described in \SecRef{ssec:physics}. The exact CG, StS, and Park's solutions have been computed with traditional numerical integrators.} In \figRef{fig:park_vs_TS1}, two different approaches are considered for Park's model, one employing the less accurate but still widely used kinetics from \refCite{Park_JTHT_2001}, derived from empirical methods or experimental data, and the other using the more recent QSS approach~\cite{Panesi_JCP_2013}, whose kinetic database is directly computed from state-to-state calculations. \blue{\FigRef{fig:park_vs_TS1} shows the evolution of the total mass fraction and internal energy content per particle of O$_2$ for the different models considered. It is evident from the figure that the coarse-grained grouping strategy employed in this work provides the closest solution to state-to-state modeling. Only \textit{Timescale 1} (or CG-DeepONet$^{\left(1,1\right)}$) of the proposed surrogate has been queried to produce the evolution of the total mass fraction of O$_2$ shown in \figRef{fig:park_vs_TS1}(a), which is in excellent agreement with the numerically-integrated CG solution.} This is because CG-DeepONet$^{\left(1,1\right)}$ implicitly contains all the information about the energy transfer processes between the 27 groups, as it has been trained with data from the integration of CGME27. However, while using only \textit{Timescale 1} is sufficient for accurately predicting the dynamics of the total mass fraction of the reactor species, the same approach may not be accurate for predicting the total internal energy content of the molecule. This is because CG-DeepONet$^{\left(1,1\right)}$ is specifically designed to model only the zeroth-order moment of the master equations and may not capture higher-order moments, such as the total internal energy content, with sufficient accuracy. Therefore, this quantity generally requires the evaluation of the overall surrogate, which includes the low-scale components CG-DeepONet$^{\left(2:4,:\right)}$. The discrepancy between the CG surrogate's predictions and the StS numerical solution in \figRef{fig:park_vs_TS1} is almost exclusively determined by the physical simplifications made by the CG model. In particular, the energy difference that can be noticed at the initial time instants is caused by the fact that the reconstructed states within each bin follow a Boltzmann distribution at the translational temperature $T$ (for the assumptions made in \SecRef{sssec:physics.framework}). In contrast, the quantum energy levels for the StS solution follow a distribution at temperature $T_{int_0}$.
\par
The proposed hierarchical architecture could be upgraded to model higher-order moments of the master equations. This improvement could involve replicating the same architecture as the CG-DeepONets to model the internal energy content of every single bin. Consequently, CG-DeepONet$^{\left(1,1\right)}$ could correctly predict both zeroth-, \textit{i.e.}, total mass, and first-order moment, \textit{i.e.}, internal energy, of O$_2$. In such a case, the low-scale components CG-DeepONet$^{\left(2:4,:\right)}$ would not be required to predict the solution shown in \figRef{fig:park_vs_TS1}(b), but they might still be necessary for providing the correct distribution function of the quantum energy states when considering other physical phenomena, such as radiation.

\subsection{Adaptive inference}\label{ssec:res.adaptive}
\begin{figure}[!htb]
	\centering
	\includegraphics[width=0.8\linewidth]{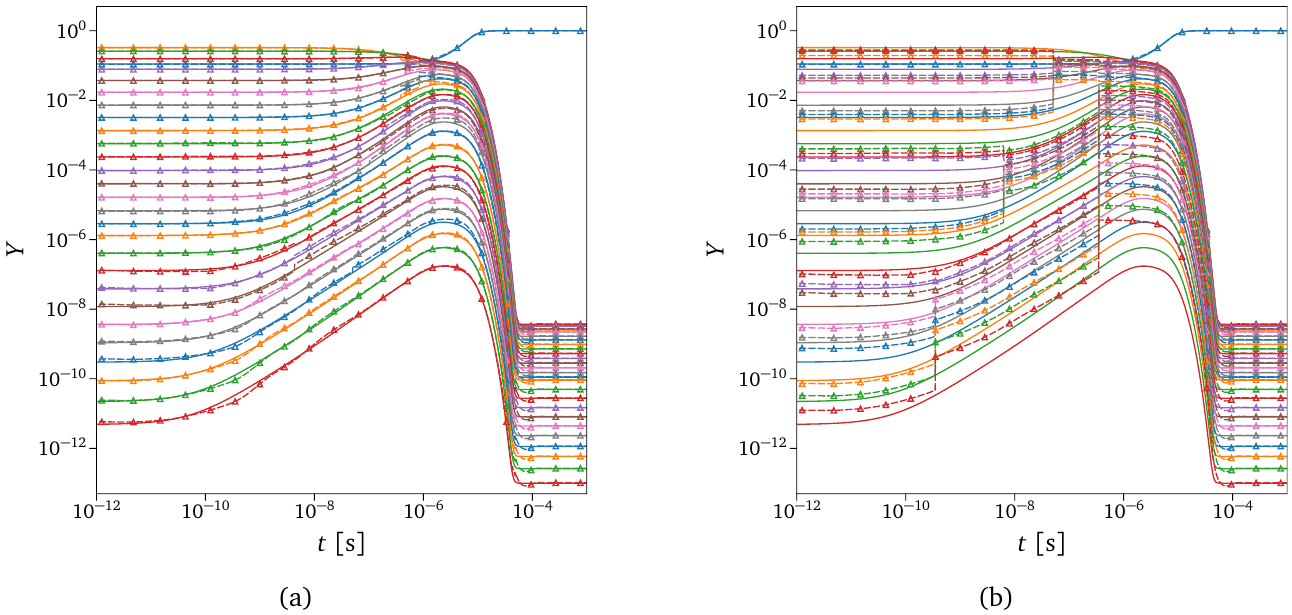}
	\caption{\textit{Exact vs.\hspace{1mm}adaptive solution for different $\delta_{\text{tol}}$.} Exact solution (solid line) versus prediction from the trained surrogate (dashed line with markers) using the adaptive technique. The isolated blue line represents the evolution of the atomic oxygen, while the remaining are the 27 groups of O$_2$ in ascending order of energy content per particle (top-down). Initial conditions used: $P_{_0}=3\,000\text{ Pa}$, $X_{\text{O}_0}=0.2$, $T_{int_0}=3\,500\text{ K}$, $T=8\,000\text{ K}$. (a) Underpredicted non-equilibrium tolerance used: $\delta_{\text{tol}}=0.1$. (b) Underpredicted non-equilibrium tolerance used: $\delta_{\text{tol}}=0.5$.}
	\label{fig:adapt_solution}
\end{figure}

The advantage of the hierarchical architecture proposed in this work is the ability to tailor the model complexity to the specific localized flow conditions to obtain a computationally efficient yet accurate physical model. Figure S4 in the Supplementary Information shows an example of the dynamics of underpredicted non-equilibrium Euclidean metric computed via \eqRef{eq:euclidean_metric} for \textit{Timescale 1} and \textit{Timescale 3} for the same test case shown in \figRef{fig:adapt_solution}. The values plotted in figure S4 can be considered a good reference for the space the proposed metric can span, as the analyzed test case exhibits considerable initial thermal and chemical non-equilibrium. It should be noted that the values of $\delta^{\left(1,1,1\right)}$ reported in figure S4(a) are almost an order of magnitude larger than figure S4(b) due to the more accurate modeling adopted in the latter. Overall, the trend is decreasing by approaching the equilibrium, except for the evident QSS region starting around 10$^{-6}$ s, where all the quantities remain constant. $\delta^{\left(3,2,6\right)}$ shows an interesting behavior in figure S4(b), which corresponds to the sixth group of the 9-groups rovibrational energy-based coarse-grained grouping strategy for \textit{Timescale 3}, the one close to the dissociation energy (5.115 eV). By observing the highly non-equilibrium StS dynamics at QSS of the states in this group (\textit{e.g.}, \figRef{fig:norm_lev_dist_comparison}), it is clear that the highest resolution possible is necessary for that region of the energy space to model the dynamics of those states accurately~\cite{Panesi_JCP_2013}.
\par
\begin{figure}[!htb]
	\centering
	\includegraphics[width=\linewidth]{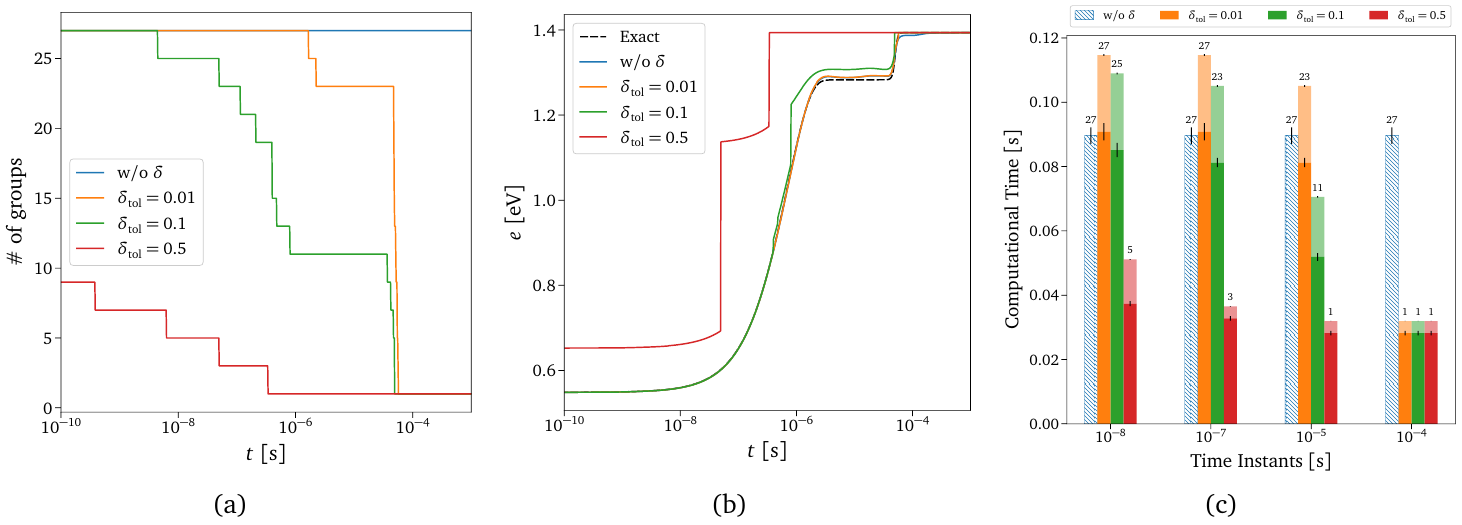}
	\caption{\textit{Adaptive solution for different $\delta_{\text{tol}}$.} The evolution of the total mass fraction of O$_2$ is not reported here, as the model accurately predicts the solution regardless of the tolerance value used. (a) Total number of actually modeled O$_2$ groups. (b) Weighted sum of the grouped-specific O$_2$ rovibrational energy, with weights given by the groups' distribution. (c) Computational cost comparison between the adaptive technique and the standalone CG-DeepONet model for single-time instant evaluation. The numbers on top of the bars are the corresponding number of groups, also shown in (a).}
	\label{fig:mul_adapt_solution}
\end{figure}

The solution obtained with the adaptive technique is compared with the exact one in \figRef{fig:adapt_solution} for two different values of the underpredicted non-equilibrium metric tolerance, $\delta_{\text{tol}}$. This value acts as a discriminant for assuming equilibrium inside each macro-group for all the timescales modeled. For $\delta_{\text{tol}}=0.1$, the adaptation starts playing effect just before the QSS region, as can also be deduced from figure S4(a), whereas for $\delta_{\text{tol}}=0.5$, it already acts at the beginning of the dynamics. We can assert that for a value of $\delta_{\text{tol}}=0.1$, the solution looks very similar to the exact one, supporting the effectiveness of the adaptive technique in terms of physical accuracy. The adaptive solutions shown in \figRef{fig:adapt_solution} have been obtained by solving the number of groups dictated by the respective $\delta$ reported in \figRef{fig:mul_adapt_solution}(a) as functions of time. From \figRef{fig:mul_adapt_solution}(a), it is evident that the number of the solved groups decreases considerably by increasing the tolerance value, confirming the validity of the proposed adaptive technique. As already demonstrated in the previous section, the prediction of the total mass fraction of O$_2$ is independent of the tolerance used since our model has been trained such that even the lowest-fidelity coarse-grained model can correctly predict the actual mass of the reactor species. However, in the case of energy, the choice of the proper tolerance can play an essential role in predicting its correct value, as shown in \figRef{fig:mul_adapt_solution}(b).
\par
\FigRef{fig:mul_adapt_solution}(c) presents a preliminary performance analysis of the adaptive technique for the different tolerance values based on a comparison with the standalone CG-DeepONet model. The reported timings are obtained as the mean of 1000 different inference evaluations of the model per each physical time instant, conducted with a single central processing unit (CPU) core. The computations shown in \FigRef{fig:mul_adapt_solution}(c) have been performed in the TensorFlow~\cite{TF_2016} environment, which means that a large part of the network evaluation time involves Python call overhead. The bar plot illustrates that the adaptive technique outperforms the standalone surrogate at later stages of the system's dynamical evolution, particularly when the composition approaches the asymptotic equilibrium value. The opaque bar chunks in \figRef{fig:mul_adapt_solution}(c) represent the contribution to the inference cost due to the Neq-DeepONets surrogate. A great advantage of this methodology is also its flexibility, as computational costs and physical accuracy can be easily balanced by tuning the tolerance value, $\delta_{\text{tol}}$. Moreover, inference with physics-informed DeepONets is trivially parallelizable with graphics processing units (GPUs), which can remarkably boost the inference timings shown in \figRef{fig:mul_adapt_solution}(c). Wang \textit{et al.}~\cite{Wang_SA_2021,WangPerdikaris_arxiv_2021} have already demonstrated that PI-DeepONets can outperform and replace conventional numerical solvers even for long-time integration.

\blue{
\subsection{One-dimensional shock case scenario}\label{ssec:1d.shock}
In this section, preliminary results of a one-dimensional numerical experiment are presented, where the constructed surrogate is tested both with and without the adaptive technique.
\par
Figures \ref{fig:1d.shock.sol}(a-b) present the final temperature and mass fraction profiles in the shock reference frame for the test case scenario described in \SecRef{sssec:shock}. In both figures, the exact solution obtained using a thermochemical library is represented with black dashed lines, while the solution obtained using the surrogate without adaptation and employing adaptive inference with tolerance values of $\delta_{\text{tol}}=0.01$ and $\delta_{\text{tol}}=0.05$ are represented by blue, orange, and green lines, respectively. The integration using the surrogate produces physically correct solutions, with the largest differences noticed at the tail of the temperature profile, in particular when the tolerance value is high. As already explained in the previous section and demonstrated in \figRef{fig:mul_adapt_solution}(b), the reason for these small discrepancies is due to the incorrect predictions of internal energy, which can result in incorrect temperature profiles while the conservation equation for total energy is integrated in time. The reconstructed microscopic distribution is also presented in \figRef{fig:1d.shock.sol}(c), showing a good agreement of the surrogate predictions with and without adaptation compared to the numerically integrated solution.
\begin{figure}[!htb]
	\centering
	\includegraphics[width=\linewidth]{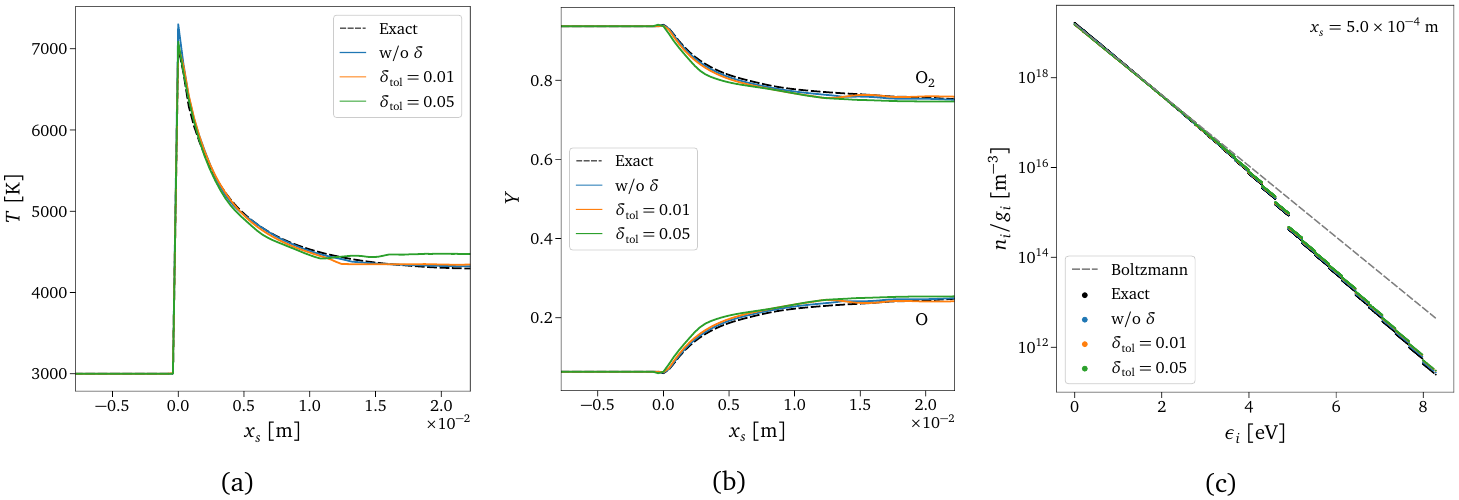}
	\caption{\textit{One-dimensional shock solution.} Comparison between exact and predicted final solutions in the shock reference frame, $x_s$. 
		\tinybullet Black dashed line: exact solution
		\tinybullet Blue line: surrogate predictions without adaptation
		\tinybullet Orange line: adaptive surrogate predictions with $\delta_{\text{tol}}=0.01$
		\tinybullet Green line: adaptive surrogate predictions with $\delta_{\text{tol}}=0.05$. (a) Translational temperature. (b) Total mass fractions of O$_2$ and O. (c) Reconstructed O$_2$ rovibrational states distribution at $x_s=5\times10^{-4}$ m and corresponding Boltzmann equilibrium distribution function.}
	\label{fig:1d.shock.sol}
\end{figure}
\par
Figures \ref{fig:1d.shock.perf}(a-b) provide a preliminary performance analysis of surrogate inference with and without adaptation. The timings are computed by evaluating only the integration time for the reactive step in \eqRef{eq:split.02} using a single CPU core within \texttt{Fortran} 2008 environment. The corresponding statistics, \textit{i.e.}, mean and standard deviation, are calculated over 500 iterations and averaged over the number of cells in the 1-$\mathrm{D}$ domain. The speedup statistics are then obtained using the formula proposed by D\'{i}az and Rubio~\cite{Diaz_Stat_2013}, which approximates the ratio of two independent normal random variables with a normal distribution. In \FigRef{fig:1d.shock.perf}(a), the speedup of the standalone surrogate is presented as a function of time step, $\Delta t$, which has been varied by changing only the number of cells and keeping everything else fixed. The surrogate inference is at least eight times faster than the serial integration performed with a conventional implicit scheme, in this case, the second-order backward differentiation formula (BDF-2). Furthermore, the maximum speedup is reached when the integration time is much longer, which is expected since the integrator may need more steps to reach the final time, unlike the surrogate inference, which is independent of the total integration time. The computed speedup depends on various factors, such as the dimension of the network, the stiffness associated with the system of equations, the scheme and tolerances used for the ODEs integration, and the length of the integrated physical time. All these details for this particular test can be found in the Supplementary Section S.3. In \FigRef{fig:1d.shock.perf}(b), a comparison is shown between the varying speedup with $\delta_{\text{tol}}$ obtained with the adaptive inference technique (light blue) and the constant one obtained with the standalone surrogate (light orange) for $\Delta t=1.33\times10^{-7}$ s. As expected, increasing the tolerance values leads to higher speedups, which is consistent with the reported timings in \figRef{fig:mul_adapt_solution}(c). However, this comes at a cost of reduced accuracy, as shown in \FigRef{fig:1d.shock.perf}(c), which presents the increasing mean relative error for temperature and total mass fraction of O$_2$ with increasing $\delta_{\text{tol}}$. The reference error values for the surrogate without adaptation are $\varepsilon_{Y_{\text{O}_2}}=0.93\%$ and $\varepsilon_T=0.58\%$. It is noteworthy that the computation of the error does not include points in the domain where the gas experiences the left or right equilibrium thermochemical states, as the surrogate predictions are not considered in those regions. The increasing error is again related to the inaccurate prediction of internal energy, as observed in the previous analysis of the temperature profile in \figRef{fig:1d.shock.sol}(a), and it may be exacerbated by the error accumulation issue, also shown by Zanardi \textit{et al.}~\cite{Zanardi_SciTech_2023} This highlights the importance of upgrading the surrogate to also model the internal energy content of each individual bin, as it can lead to improved accuracy in terms of the macroscopic quantities of interest. Nevertheless, this approach holds promise when scaled to multi-dimensional CFD simulations with millions of unknowns. For example, in hypersonic simulations, most domain points may lie in the equilibrium or near-equilibrium regions, while only a few points may be in strong non-equilibrium regions (such as shock proximity) where the evaluation of the entire surrogate is needed. In light of these considerations and the performance analysis performed, the adaptive technique has the potential to outperform the standalone model in a multi-dimensional simulation framework.
\begin{figure}[!htb]
	\centering
	\includegraphics[width=\linewidth]{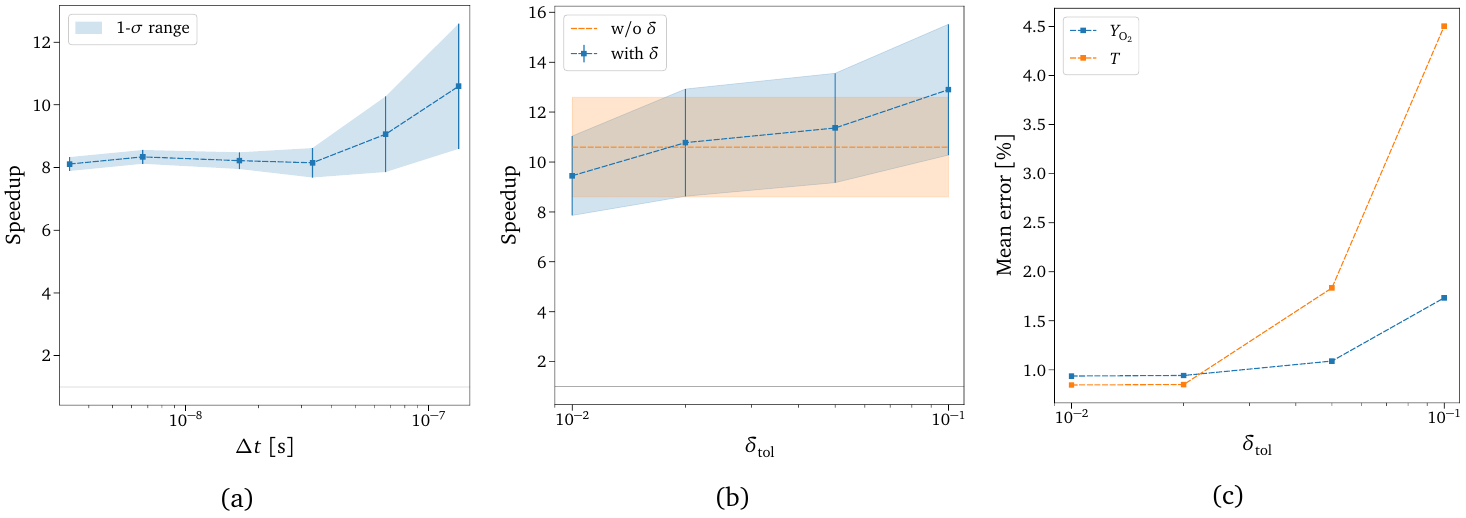}
	\caption{\textit{Surrogate performances and accuracy in one-dimensional simulation.} (a) Speedup achieved using the surrogate without adaptation with increasing $\Delta t$, which is equivalent to reducing the number of cells while maintaining a constant CFL. (b) Speedup achieved using adaptive surrogate inference with increasing $\delta_{\text{tol}}$. (c) Mean relative percentage error for $Y_{\text{O}_2}$ and $T$ as a function of $\delta_{\text{tol}}$. The constant reference values for the surrogate without adaptation are $\varepsilon_{Y_{\text{O}_2}}=0.93\%$ and $\varepsilon_T=0.58\%$.}
	\label{fig:1d.shock.perf}
\end{figure}
}

\section*{Conclusions}\label{sec:concl}We proposed a new machine learning-based paradigm inspired and constrained by physical laws for solving multiscale non-equilibrium flows. The designed model (CG-DeepONet) sequentially learned the integral solution operator for multi-fidelity coarse-grained master equations by employing a physics-inspired hierarchical architecture, where physics-informed DeepONet (PI-DeepONet) represents the core element. Furthermore, we developed a controller-acting surrogate (Neq-DeepONet) to learn the dynamics of the underpredicted degree of non-equilibrium to tailor the model's accuracy to the local non-equilibrium conditions. \violet{Finally, by combining the two, we designed a novel adaptive pruning inference technique for non-equilibrium thermochemical processes, which showed flexibility in balancing accuracy and computational cost.
	\par
	Overall, the proposed framework incorporates different key elements that enforce the underlying physics into the surrogate:
	\begin{inparaitem}
		\item[i)] the physics-based dimensionality reduction in the state space;
		\item[ii)] the additional layers enforcing the Boltzmann distribution functions, which in turn allow the imposition of prior distributions for the network parameters. When propagated to the state populations (\textit{e.g.}, mass fractions), such priors provide physically consistent solutions even when the surrogate is not trained (\textit{i.e.}, equilibrium distributions);
		\item[iii)] the physics-informed loss;
		\item[iv)] the hierarchical architecture and the related sequential fine-tuning transfer learning between different time scales, with mass conservation enforced;
		\item[v)] the online pruning of the surrogate at the prediction phase through a parsimony-based approach that relies on an additional controller-acting surrogate informed by a non-equilibrium variable.
\end{inparaitem}}
\par
The methodology was applied to the study of chemical kinetics relevant for application to hypersonic flight and was tested on oxygen mixtures. However, the framework is not constrained to the chosen thermochemical configuration, but it can be extended to Air-5 mixtures (\textit{i.e.}, simultaneously with N$_2$, O$_2$, NO, N, and O species) or even other fields of physics spanning a wide range of temporal scales, such as electromagnetism, magnetohydrodynamics, and more generally, plasma physics. \violet{The proposed framework was tested in 0-$\mathrm{D}$ and 1-$\mathrm{D}$ configurations, and the following results were obtained:
\begin{itemize}[-]
	\item In 0-$\mathrm{D}$ scenarios, the CG-DeepONet surrogate alone showed excellent physical accuracy compared to the numerical integration of the master equation, with a maximum relative error of 4.5\%. It also exhibited good computational efficiency when the adaptive method was used, gaining more than 3X speedup in the regions of weak non-equilibrium.
	\item The 1-$\mathrm{D}$ numerical experiment demonstrated the flexibility of the proposed method in capturing complex dynamics and confirmed the good performances and accuracy of both standalone and adaptive versions of the constructed surrogate. The relative error was in the range of 1\%-4.5\% with a corresponding 8X-13X speedup compared to conventional implicit schemes employed in an operator-splitting integration framework. As expected, the choice of high tolerances for the adaptive schemes and the consequent lack of degrees of freedom in characterizing the rovibrational distribution generated error accumulations in the predictions of the overall O$_2$ internal energy. In future work, we will treat the group temperatures as state variables together with the species mass fractions. This addition will have two benefits. Firstly, it will allow us to achieve comparable accuracy with fewer groups. Secondly, it will enable the accurate prediction of the O$_2$ internal energy by relying only on the first scale (\textit{i.e.}, CG-DeepONet$^{\left(1,1\right)}$), similar to what was achieved for the mass fractions (\textit{e.g.}, \figRef{fig:park_vs_TS1}).
\end{itemize}
\par
Future work will extend and test the framework to 2-$\mathrm{D}$ and 3-$\mathrm{D}$ simulations, leveraging its ability to be designed and constructed independently of geometric features of the problem. Additionally, alternative neural operator approaches other than DeepONets will be explored to mitigate the issue of error accumulation. Beyond the application and the numerical outcomes, this work serves as an example on how physics and machine learning can enhance each other, aiming for more interpretable and robust ML-based tools for the scientific community.}

\section*{Data availability}
The dataset used in the current study is available from the corresponding author upon reasonable request.

\section*{Code availability}
The code used in the current study is available from the corresponding author upon reasonable request.

\bibliography{sections/biblio}

\section*{Acknowledgements}
The work is supported by the Vannevar Bush Faculty Fellowship OUSD(RE) Grant No: N00014-21-1-295 with Prof. Marco Panesi as the Principal Investigator. The authors wish to thank Dr. Pietro Novelli (Istituto Italiano di Tecnologia, Italy) for many helpful discussions. The views and conclusions contained herein are those of the authors and should not be interpreted as necessarily representing the official policies or endorsements, either expressed or implied, of the U.S. government.

\section*{Author contributions}
I.Z. and S.V. conceptualization and methodology; I.Z. software; M.P. supervision and funding acquisition. All authors participated in the data analysis, paper writing, and manuscript revision.

\section*{Competing interests}
The authors declare no competing interests.

\section*{Additional information}
Correspondence and requests for materials should be addressed to M.P.

\clearpage
\input{sections/suppl}

\end{document}

%% file: sections/suppl.tex
\graphicspath{{./figs/suppl/}}

\renewcommand*{\thetable}{S\arabic{table}}
\renewcommand*{\thefigure}{S\arabic{figure}}
\renewcommand*{\theequation}{S\arabic{equation}}
\renewcommand*{\thesection}{S.\arabic{section}}
\renewcommand*{\thesubsection}{\thesection.\arabic{subsection}}
\renewcommand*{\thesubsubsection}{\thesubsection.\arabic{subsubsection}}
\renewcommand*{\thealgocf}{S\arabic{algocf}}

{\raggedright\sffamily\bfseries\fontsize{20}{25}\selectfont Supplementary Information \par}%

\section{Physical modeling}\label{sec:suppl:physics}
The gaseous mixtures considered in the proposed framework consist solely of oxygen atoms and molecules, both assumed to be in their ground electronic states. The set of pseudo-species, $i$, is defined as $\mathcal{I}=\left\{\text{O},\text{O}_2^{\left(i\right)}\right\}$, encompassing all the possible internal energy degrees of freedom of the system.

\subsection{Thermodynamics}\label{ssec:suppl.thermo}
The gas pressure follows from Dalton's law,
\begin{equation}
	p=\sum_{i\in\mathcal{I}}n_i k_\mathrm{B} T\hspace{2mm},
\end{equation}
where $n_i$ stands for the number density of the pseudo-species $i$, whereas $k_\mathrm{B}$  denotes Boltzmann’s constant. The gas density reads $\rho=\sum_{i\in\mathcal{I}}\rho_i$, where the partial densities are related to the number densities via $\rho_i=m_in_i$, with $m_i$ being the (particle) mass of $i$. The energy per unit-mass of the individual pseudo-species may be written as
\begin{equation}\label{eq:energy}
	e_i=e_{i,\text{tr}}+e_{i,\text{int}}+\Delta h_{i,\text{f}}\hspace{2mm},
\end{equation}
where translational contribution follows from the principle of equipartition of energy~\cite{Callen_AJP_1998}:
\begin{equation}\label{eq:trasl.en}
	e_{i,\text{tr}}=\frac{3}{2}\frac{k_\mathrm{B}}{m_i}T\hspace{2mm}.
\end{equation}
The symbol $\Delta h_{i,\text{f}}$ in \eqRef{eq:energy} denotes the formation enthalpy at 0 K, whereas the remaining term, $e_{i,\text{int}}$, accounts for the internal energy degree of freedom of the pseudo-species $i$:
\begin{enumerate}[i.]
	\item If the state-to-state (StS) modeling is used, the term $e_{i,\text{int}}$ represents a particular rovibrational energy state, denoted as $\epsilon_i$, where $i=(s,v,J)$ with $s$ representing the species, $v$ representing the vibrational quantum number, and $J$ representing the rotational quantum number.
	\item If coarse-grained modeling is used, $\rho_i$ indicates the density of a group of states and $e_{i,\text{int}}$ can be expressed as follows:
	\begin{equation}\label{eq:CG_energy}
		e_{i,\text{int}}=e_P=\sum_{i\in\mathcal{I}_P}\frac{q_i\left(T_P\right)}{Q_P\left(T_P\right)}\epsilon_i\hspace{2mm},
	\end{equation}
	with the additional new terms appearing in \eqRef{eq:CG_energy} described in \SecRef{ssec:suppl.kin}.
	\item If multi-temperature (MT) models~\cite{Park_Book_1990} are employed, $e_{i,\text{int}}$ accounts for the energy of \emph{thermalized} internal degrees of freedom (\emph{e.g.}, rotation, vibration). For a conventional two-temperature (2T) formulation~\cite{Park_Book_1990,Park_JTHT_1993}, which is a particular class of MT models, the expression of $e_{i,\text{int}}$ for a diatomic molecule described by the rigid-rotor and harmonic oscillator models is~\cite{VincentiKruger_Book_1965,Anderson_Book_2019}:
	\begin{equation}
		e_{i,\text{int}}=e_{i,\text{r}}\left(T_\text{r}\right)
			+e_{i,\text{v}}\left(T_\text{v}\right)\hspace{2mm},
	\end{equation}
	with
	\begin{align}
		e_{i,\text{r}}\left(T_\text{r}\right)&
		=\frac{k_\mathrm{B}}{m_i}T_\text{r}\hspace{2mm},\\ e_{i,\text{v}}\left(T_\text{v}\right)&
		=\frac{k_\mathrm{B}}{m_s}\frac{\theta_s^\text{v}}{\exp\left(\theta_s^\text{v}/T_\text{v}\right)-1}\hspace{2mm},
	\end{align}
	where $T_\text{r}$ and $T_\text{v}$ are, respectively, the rotational and vibrational temperatures, whereas $\theta_s^\text{v}$ is the characteristic vibrational temperature. In Park's two-temperature model, the fast equilibration between rotational and translational energy mode is assumed (\emph{i.e.}, $T_\text{r}=T$).
\end{enumerate}
Collecting the above formulae, the energy per unit-mass of the gas as a whole can be defined as follows:
\begin{equation}
	e=\sum_{i\in\mathcal{I}}Y_i\left(e_{i,\text{tr}}+e_{i,\text{int}}+\Delta h_{i,\text{f}}\right)\hspace{2mm},
\end{equation}
where the mass fractions are $Y_i=\rho_i/\rho$.

\subsection{Coarse-grained modeling}\label{ssec:suppl.kin}
This work employs a log-linear form of the distribution function to represent the population within each individual bin, which results in a thermalized local Boltzmann distribution defined as follows:
\begin{equation}
	\mathcal{F}_P^i\left(\epsilon_i\right):\quad
	\log\left(\frac{g_i}{n_i}\right)=\alpha_P+\beta_P\epsilon_i\hspace{2mm}.
\end{equation}
The bin-specific coefficients $\alpha_P$ and $\beta_P$ are formulated in terms of the macroscopic constraints, total bin population $n_P$ and energy $e_P$,
\begin{equation}
	n_P=\sum_{i\in\mathcal{I}_P}n_i\hspace{1mm},
	\quad
	e_P=\sum_{i\in\mathcal{I}_P}n_i\epsilon_i\hspace{2mm},
\end{equation}
where $\mathcal{I}_P$ indicates the set of rovibrational states contained in the $P$-th group. The bin internal temperature $T_P$ can be used instead of $\beta_P$ to characterize the bin distribution function,
\begin{equation}%
	\beta_P=\frac{1}{k_BT_P}\hspace{2mm},
\end{equation}
while the coefficient $\alpha_P$ can then be defined as follows:
\begin{equation}
	\alpha_P=\log\left(\frac{Q_P}{n_P}\right)\hspace{2mm},
\end{equation}
where $Q_P$ is the group internal partition function,
\begin{equation}\label{eq:group_part_fn}
	Q_P\left(T_P\right)=\sum\limits_{i\in\mathcal{I}_P}q_i\left(T_P\right)\hspace{2mm},
\end{equation}
with
\begin{equation}\label{eq:state_part_fn}
	q_i\left(T_P\right)=g^\text{e}g_i\exp\left(-\frac{\epsilon_i}{k_BT_P}\right)
\end{equation}
being the $i$-th level contribution, $k_B$ the Boltzmann’s constant, and $g^\text{e}$ the degeneracy of the electronic ground state.

\subsubsection{Zero-dimensional macroscopic equations}
Considering O$_2+$O system, as the group temperatures $T_P$ are assumed to be equal to the translational temperature $T$, only the zeroth-order moment of the StS master equations~\cite{Panesi_JCP_2013, Panesi_PR_2014} is required to model the reactor dynamics:
\begin{equation}
	\begin{dcases}
		\frac{dn_P}{dt}=\Omega^0_P=-\sum_{\mathcal{I}_Q\in\mathcal{I}_{\text{O}_2}}K^E_{PQ}n_Pn_\text{O}
		+\sum_{\mathcal{I}_Q\in\mathcal{I}_{\text{O}_2}}K^E_{QP}n_{_Q}n_\text{O}-K^D_Pn_Pn_\text{O}
		+K^R_Pn_\text{O}^3\qquad\forall\hspace{1mm}\mathcal{I}_P\in\mathcal{I}_{\text{O}_2}\\
		\frac{dn_\text{O}}{dt}=\Omega^0_\text{O}=
		\sum_{\mathcal{I}_P\in\mathcal{I}_{\text{O}_2}}K^D_Pn_Pn_\text{O}
		-\sum_{\mathcal{I}_P\in\mathcal{I}_{\text{O}_2}}K^R_Pn_\text{O}^3
	\end{dcases}\hspace{2mm},
\end{equation}
with $\mathcal{I}_{\text{O}_2}$ being the set of pseudo-species, \textit{i.e.}, groups,  of O$_2$. The group-specific rate coefficients, $K^E_{PQ}$ and $K^D_P$, are obtained from the state-specific ones, $k^E_{ij}$ and $k_i^D$, as a weighted average based on the Boltzmann distribution function over $\mathcal{I}_P$:
\begin{align}\label{eq:Kgrouped}
	K^E_{PQ}\left(T,T_P\right)&=\sum_{i\in\mathcal{I}_P}\sum_{j\in\mathcal{I}_Q}
	\frac{q_i\left(T_P\right)}{Q_P\left(T_P\right)}k^E_{ij}\left(T\right)\hspace{2mm},\\
	K^D_P\left(T,T_P\right)&=\sum_{i\in\mathcal{I}_P}
	\frac{q_i\left(T_P\right)}{Q_P\left(T_P\right)}k_i^D\left(T\right)\hspace{2mm}.
\end{align}

\clearpage
\section{Neural operators}\label{sec:suppl:ml}

\subsection{DeepONet}\label{ssec:suppl:ml.opnn}
\subsubsection{Vanilla architecture}\label{sssec:suppl:ml.vanilla.opnn}
The vanilla version of the DeepONet consists of one branch net and one trunk net. To account for the problem's multi-dimensionality, the feature embedding $\bm{\alpha}$ (and equivalently $\bm{\phi}$) has a dimension of $p \times D$, where $p$ is the number of modes (in a POD sense) and $D$ is the number of output variables. To ensure a continuous and differentiable representation of the output functions, the branch and trunk network outputs are split into $D$ $p$-dimensional vectors, which are merged together via dot product as follows:
\begin{equation}
	\widehat{G}^{(i)}(\bm{u})(\bm{y})=\sum_{k=(i-1)p+1}^{ip} \alpha_{k}(\bm{u})\phi_{k}(\bm{y})
	\qquad\text{for}\hspace{2mm}i=1,\dots,D\hspace{2mm}.
	\label{eq:vanilla_dot_prod}
\end{equation}
To ensure a fair comparison, the vanilla DeepONet, summarized in \tabRef{table:vanilla_arch}, has been designed to have almost the same number of parameters (231 388) as the CG-DeepONets described in \SecRef{ssec:suppl.network}, and it has been trained and tested on the same datasets. The entirely optimization has been performed under identical conditions in terms of hyper-parameters, including number of epochs, optimizer type, learning rate, and regularization.
\begin{table}[!htb]
	\centering
	\begin{tabular}{cccc}
		\toprule
		Sub-networks & Type & Layers Width & $\sigma$ \\
		\midrule
		Branch & FNN & [240, 240, 224] & tanh$\times$2 + linear \\
		Trunk & FNN & [240, 240, 224] & tanh$\times$2 + linear \\
		\bottomrule
	\end{tabular}
	\caption{\textit{Vanilla DeepONet architecture.} FNN is the conventional feed-forward neural network, and $\sigma$ are the activation functions.}
	\label{table:vanilla_arch}
\end{table}
\TabRef{table:vanilla_error} presents the four largest errors of the inferred solution using the vanilla DeepONet, following the same procedure described in the \ManuSecRef{Accuracy}. In the table, the apex refers to the O$_2$ group. The results demonstrate that although the errors are within acceptable limits, they are nearly double compared to the ones reported in \TabRef{table:error} in the manuscript, indicating that the vanilla DeepONet does not perform as well as the CG-DeepONets on this particular problem. Furthermore, the vanilla DeepONet cannot perform adaptive inference, which is essential for achieving increased speedup in the prediction phase.
\begin{table}[!htb]
	\centering
	\renewcommand{\arraystretch}{1.6}
	\begin{tabular}{c|c}
		\toprule
		Group & Rel. error [\%] \\
		\midrule
		$\widehat{Y}^{(27)}$ & $6.80\pm4.61$ \\
		$\widehat{Y}^{(26)}$ & $6.55\pm4.39$ \\
		$\widehat{Y}^{(24)}$ & $6.43\pm4.12$ \\
		$\widehat{Y}^{(23)}$ & $6.42\pm3.88$ \\
		\bottomrule
	\end{tabular}
	\caption{\textit{Vanilla DeepONet test error.} The four highest mean relative $L^2$-norm testing errors (with standard deviations) of the trained vanilla DeepONet.}
	\label{table:vanilla_error}
\end{table}

\subsubsection{Shared-trunk architecture}\label{sssec:suppl:ml.shared.trunk.opnn}
\begin{figure}[!htb]
	\centering
	\includegraphics[width=0.5\linewidth]{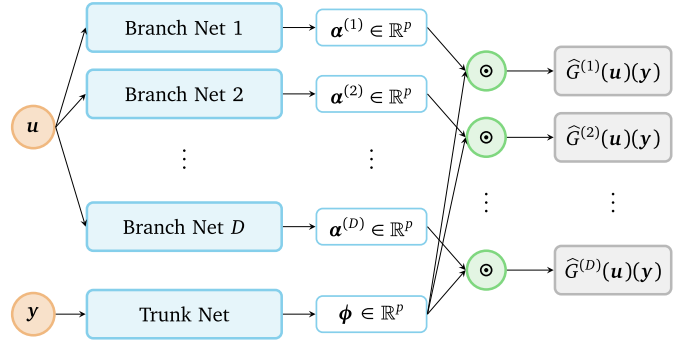}
	\vspace{1mm}
	\caption{\textit{Multi-output DeepONet.} The modified architecture consists of multiple ``branch nets'' (one for each output) for extracting latent representations of the input functions and one ``trunk net'' for extracting latent representations of the input coordinates at which the output functions are evaluated.}
	\label{fig:shared.trunk.opnn}
\end{figure}

As depicted in \FigRef{fig:shared.trunk.opnn}, the shared-trunk architecture is characterized by multiple branches, each corresponding to an output variable and a shared trunk network. This design allows for computational efficiency, as the shared trunk can be used for multiple output variables. However, it is effective only when the dynamics of the modeled variables are similar enough to share the same basis, as discussed in \refCite{Venturi_CMAME_2023}.

\subsection{Multi-scale hierarchical coarse-grained model}\label{ssec:suppl.network}

\subsubsection{Hyper-parameter settings}\label{sssec:suppl.hyper}
\TabRef{table:archit_cg} summarizes the CG-DeepONets architecture, where a modified version of the DeepONet proposed by Wang \textit{et al.} [Eqs. (3.23)-(3.29) in \refCite{Wang_JSC_2022}] is used. The network is trained via mini-batch stochastic gradient descent for $10^4$ iterations using the Adam optimizer for each step (a-d) described in the \ManuSecRef{Training Strategy}. The last step using physics-informed optimization techniques has been performed for $5\times10^3$ epochs. For each training step, to obtain a set of training and validation data, $2\times N$ initial conditions have been sampled using the Latin Hypercube strategy with $N=2\,048$. Half of them have been selected as training scenarios using the stratified sampling method, and the remaining half as validation. For each $i$-th initial condition, $P=128$ and $P=32$ data points for training and validation have been log-uniformly sampled in time. To generate the test data set, we randomly sampled 100 unseen initial conditions and obtained the corresponding numerical solutions by integrating the ODE using a conventional numerical integrator.
\begin{table}[!htb]
	\centering
	\begin{tabular}{ccccccc}
		\toprule
		& \multirow{2}{*}{\# DeepONets} & \multicolumn{5}{c}{Single DeepONet} \\
		\cmidrule(lr){3-7}
		& & Sub-networks & Type & $p$ & Layers Width & $\sigma$ \\
		\midrule
		\multirow{2}{*}{\textit{Timescale 1}} & \multirow{2}{*}{1} & 2 Branches & ResFNN & \multirow{2}{*}{8} & [32, 32, $p$] & tanh$\times$3 \\
		& & 1 Trunk & ResFNN & & [32, 32, $p$] & tanh$\times$3 \\
		\midrule
		\multirow{2}{*}{\textit{Timescale 2}} & \multirow{2}{*}{1} & 3 Branches & ResFNN & \multirow{2}{*}{16} & [48, 48, $p$] & tanh$\times$3 \\
		& & 1 Trunk & ResFNN & & [48, 48, $p$] & tanh$\times$3 \\
		\midrule
		\multirow{2}{*}{\textit{Timescale 3}} & \multirow{2}{*}{3} & 3 Branches & ResFNN & \multirow{2}{*}{16} & [48, 48, $p$] & tanh$\times$3 \\
		& & 1 Trunk & ResFNN & & [48, 48, $p$] & tanh$\times$3 \\
		\midrule
		\multirow{2}{*}{\textit{Timescale 4}} & \multirow{2}{*}{9} & 3 Branches & ResFNN & \multirow{2}{*}{16} & [48, 48, $p$] & tanh$\times$3 \\
		& & 1 Trunk & ResFNN & & [48, 48, $p$] & tanh$\times$3 \\
		\bottomrule
	\end{tabular}
	\caption{\textit{CG-DeepONets architecture.} $p$ is the dimension of the features embedding, ResFNN is a novel neural network architecture proposed by Wang \textit{et al.} [Eqs. (2.33)-(2.37) in \refCite{Wang_arXiv_2020}], and $\sigma$ are the activation functions.}
	\label{table:archit_cg}
\end{table}
\par
It has to be mentioned that an input transformation layer is used to modify the input features. For the trunk net, the time $t$ has been linearly scaled by a factor of $10^7$, while for the branch net, the temperature $T$ has been normalized between 0 and 1. The total number of parameters of the network is 230$\,$106.

\subsubsection{Extra test cases}\label{sssec:suppl.tests}
\begin{figure}[!htb]
	\centering
	\includegraphics[width=0.8\textwidth]{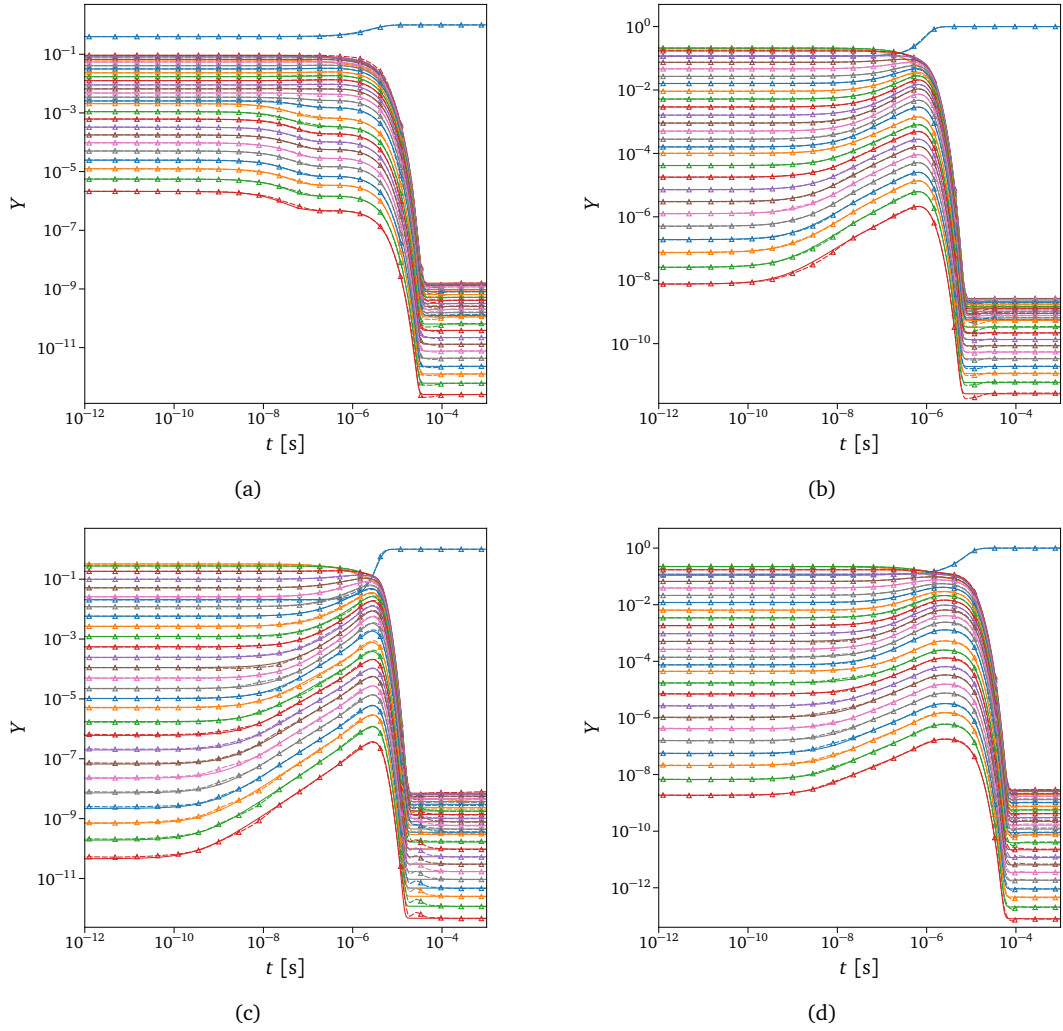}
	\caption{\textit{Extra test cases for CG-DeepONets.}}
	\label{fig:extra_tests}
\end{figure}

\FigRef{fig:extra_tests} shows extra testing cases, similarly to what has been shown in the \ManuSecRef{Inference}.

\subsubsection{Loss histories}\label{sssec:suppl.losses}
\begin{figure}[!htb]
	\centering
	\includegraphics[width=\textwidth]{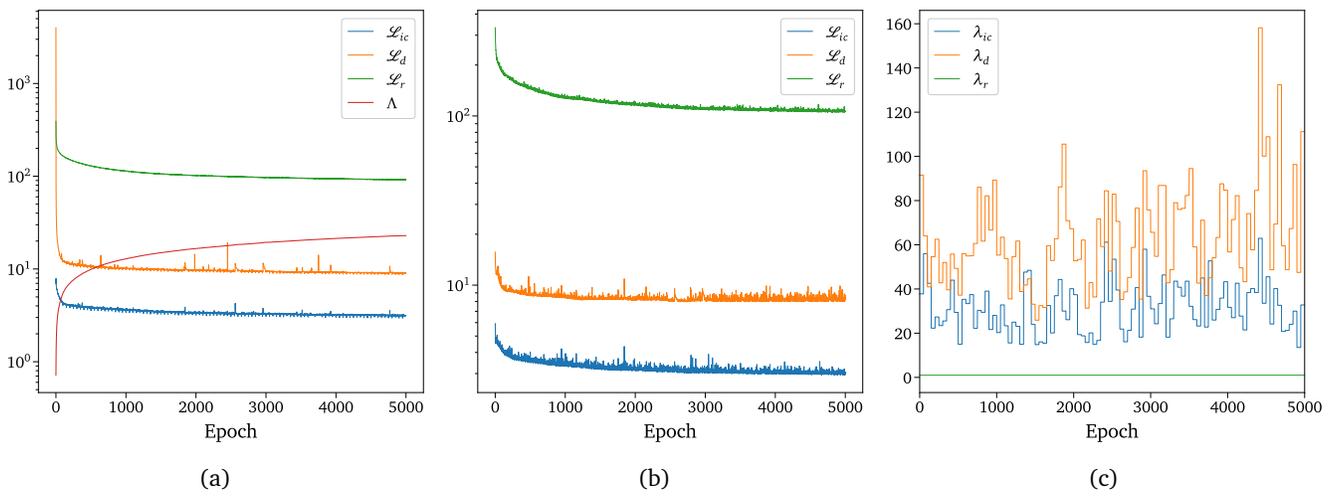}
	\caption{\textit{Loss histories of the last physics-informed training step.} (a) Training losses. (b) Validation losses. (c) Loss weighting coefficients.}
	\label{fig:losses}
\end{figure}

In this section, only the last and most interesting hybrid training step of the strategy described in the \ManuSecRef{Training Strategy} is presented. Figures \ref{fig:losses}(a-b) show the training and validation losses, respectively, while \FigRef{fig:losses}(c) shows the evolution of the weights coefficients $\lambda_i$, automatically tuned every 50 iterations with the learning rate annealing technique described by Alg. 2.1 in \refCite{Wang_arXiv_2020}. $\lambda_r$ is fixed and equal to 1 since the ODE residual loss, $\mathcal{L}_{r}$, is used as the reference value for computing $\lambda_d$ and $\lambda_{ic}$, while the hyper-parameter $\alpha$ of the tuning procedure has been set to 0.7.

\newpage
\subsection{Adaptive inference}\label{ssec:suppl.adapt}
\subsubsection{Hyper-parameter settings}\label{sssec:suppl.adapt.hyper}
\TabRef{table:archit_neq} summarizes the Neq-DeepONets architecture. We trained the model by employing a similar strategy used for the CG-DeepONets, except that all the timescales have been trained simultaneously in this case. The version of the DeepONet used to construct the controller-acting surrogate is an augmented version called \textit{flexDeepONet} proposed by Venturi and Casey [figure (8) in \refCite{Venturi_CMAME_2023}]. The network is trained via mini-batch stochastic gradient descent for $2\times10^4$ iterations using the Adam optimizer and the mean absolute percentage error as the loss function.
\begin{table}[!htb]
	\centering
	\begin{tabular}{ccccccc}
		\toprule
		& \multirow{2}{*}{\# DeepONets} & \multicolumn{5}{c}{Single DeepONet} \\
		\cmidrule(lr){3-7}
		& & Sub-networks & Type & $p$ & Layers Width & $\sigma$ \\
		\midrule
		\multirow{3}{*}{\textit{Timescale 1}} & \multirow{3}{*}{1} & 1 Branch & FNN & \multirow{2}{*}{8} & [32, 32, $p$] & tanh$\times$2 + linear \\
		& & 1 Trunk & FNN & & [32, 32, $p$] & tanh$\times$2 + linear \\
		& & 1 PreNet & FNN & - & [16, 16, 2] & tanh$\times$2 + linear \\
		\midrule
		\multirow{3}{*}{\textit{Timescale 2}} & \multirow{3}{*}{1} & 3 Branches & FNN & \multirow{2}{*}{16} & [48, 48, $p$] & tanh$\times$2 + linear \\
		& & 1 Trunk & FNN & & [48, 48, $p$] & tanh$\times$2 + linear \\
		& & 1 PreNet & FNN & - & [16, 16, 2] & tanh$\times$2 + linear \\
		\midrule
		\multirow{3}{*}{\textit{Timescale 3}} & \multirow{3}{*}{3} & 3 Branches & FNN & \multirow{2}{*}{16} & [48, 48, $p$] & tanh$\times$2 + linear \\
		& & 1 Trunk & FNN & & [48, 48, $p$] & tanh$\times$2 + linear \\
		& & 1 PreNet & FNN & - & [16, 16, 2] & tanh$\times$2 + linear \\
		\bottomrule
	\end{tabular}
	\caption{\textit{Neq-DeepONets architecture.} $p$ is the dimension of the features embedding, FNN is the conventional feed-forward neural network, and $\sigma$ are the activation functions.}
	\label{table:archit_neq}
\end{table}
How concerns the input transformation layer, for the trunk net, the time $t$ has been linearly scaled by a factor of $10^7$ and then log-transformed, while for the branch net, the temperature $T$ has been normalized between 0 and 1. An exponential transformation function is applied to the output of each DeepONet. The total number of parameters of the network is 75$\,$487.
\begin{figure}[!htb]
	\centering
	\includegraphics[width=0.8\textwidth]{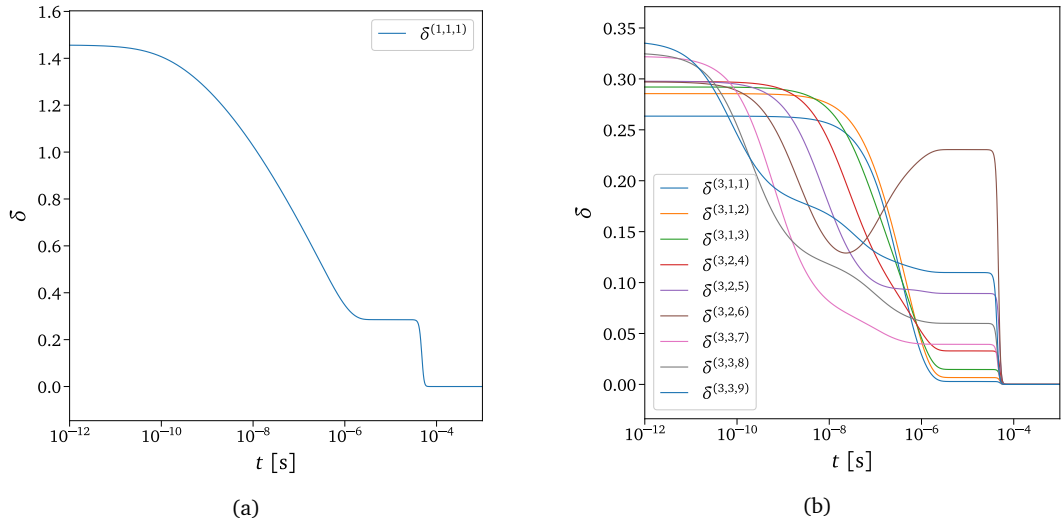}
	\caption{\textit{Underpredicted non-equilibrium metrics.} Dynamics of underpredicted non-equilibrium metrics evaluated by reconstructing the 27 groups with predictions from \textit{Timescale 1} (a) and \textit{Timescale 3} (b) for the same test case shown in \figRef{fig:adapt_solution}.}
	\label{fig:adapt_metrics}
\end{figure}

\subsubsection{Inference solution and accuracy}\label{sssec:suppl.adapt.infAcc}
In \figRef{fig:adapt_inferred}, a comparison between the exact and inferred solutions of the trained model is presented for different test cases, while \tabRef{table:adapt_error} reports the four highest errors of the inferred solution, similar to what has been done in Section the \ManuSecRef{Accuracy}.
\begin{table}[!htb]
	\centering
	\renewcommand{\arraystretch}{1.6}
	\begin{tabular}{c|c}
		\toprule
		Neq. metric & Rel. error [\%] \\
		\midrule
		$\delta^{\left(3,2,6\right)}$ & $1.81\pm1.50$ \\
		$\delta^{\left(3,1,3\right)}$ & $1.48\pm1.34$ \\
		$\delta^{\left(2,1,2\right)}$ & $1.46\pm1.25$ \\
		$\delta^{\left(3,2,5\right)}$ & $1.44\pm1.24$ \\
		\bottomrule
	\end{tabular}
	\caption{\textit{Neq-DeepONets test error.} The four highest mean relative $L^2$-norm testing errors (with standard deviations) of the trained Neq-DeepONets surrogate.}
	\label{table:adapt_error}
\end{table}
\begin{figure}[!htb]
	\centering
	\includegraphics[width=0.92\textwidth]{05.pdf}
	\caption{\textit{Test cases for Neq-DeepONets.} Case (a) is the one shown in \figRef{fig:adapt_solution}.}
	\label{fig:adapt_inferred}
\end{figure}

\subsubsection{Adaptive inference algorithm}\label{sssec:suppl.adapt.algo}
\AlgRef{alg:adapt} presents in detail all the steps of the adaptive technique used to get the inferred solutions.
\input{sections/extras/adapt_algo}

\clearpage
\section{One-dimensional numerical experiment}\label{sec:suppl:shock}
In this section, the construction of the surrogates used in the one-dimensional numerical experiment is described, which involves the following steps:
\begin{enumerate}[i.]
	\item Running the exact solution using the computational framework described in \SecRef{ssec:shock.comp_frame} and the configuration described in the \ManuSecRef{One-dimensional numerical experiment}.
	\item Collecting all the possible thermochemical states experienced by the gas in the 1-$\mathrm{D}$ simulation and fitting a 29-dimensional multivariate Gaussian-based kernel density estimator (KDE) to the data, which includes temperature, $T$, and densities $\rho_i$ of O and the 27 groups of O$_2$.
	\item Sampling $N=5120$ initial thermochemical states from the constructed KDE for training and validation, and using $N=100$ states for testing. Then, performing 0-$\mathrm{D}$ simulations for all the sampled initial states.
	\item Conducting a singular value decomposition (SVD) analysis on the trajectories obtained from the previous step to estimate the number of modes required for modeling each timescale in the CG-DeepONets surrogate~\cite{Venturi_CMAME_2023}. Similarly, utilizing \EqRef{eq:euclidean_metric} to obtain data for Neq-DeepONets surrogate from the generated trajectories, and performing the same SVD analysis.
	\item Constructing the datasets for CG-DeepONets and Neq-DeepONets by sampling 72 points for training and 18 points for validation from the previously generated trajectories in a time window of [10$^{-10}$,10$^{-6}$] s, which encompasses the time steps used in the numerical experiment.
	\item Training both models, CG-DeepONets and Neq-DeepONets, following the procedures described in \SecRef{sssec:suppl.hyper} and \SecRef{sssec:suppl.adapt.hyper}, respectively. However, in this case, the hybrid step described in the \ManuSecRef{Training Strategy} is not performed.
\end{enumerate}

\subsection{Surrogate hyper-parameter settings}\label{ssec:shock.network}
The architectures of CG-DeepONets and Neq-DeepONets employed for the 1-$\mathrm{D}$ test case are summarized in \TabRef{table:archit_cg.1d} and \TabRef{table:archit_neq.1d}, respectively. Each trunk of the Neq-DeepONets has been fitted with a radial basis function (RBF) interpolator after training to accelerate the network evaluation. The version of DeepONet used to construct the CG-DeepONets surrogate is the \textit{flexDeepONet} proposed by Venturi and Casey~\cite{Venturi_CMAME_2023}. In this case, a unique global \textit{PreNet} for each modeled \textit{Timescale} is used, constructed with a feedforward neural network (FNN) architecture consisting of layers with widths [16, 16, 2] and activation functions [\textit{tanh, tanh, linear}]. Both initial conditions and time inputs are log-transformed in both surrogates. Additionally, an exponential transformation function is applied to the output of each DeepONet in the Neq-DeepONets surrogate, as well as to the one modeling the temperature in the CG-DeepONets surrogate.
\begin{table}[!htb]
	\centering
	\begin{tabular}{cccccccc}
		\toprule
		& \multirow{2}{*}{\# PreNets} & \multirow{2}{*}{\# DeepONets} & \multicolumn{5}{c}{Single DeepONet} \\
		\cmidrule(lr){4-8}
		& & & Sub-networks & Type & $p$ & Layers Width & $\sigma$ \\
		\midrule
		\multirow{2}{*}{\textit{Temperature}} & \multirow{2}{*}{1} & \multirow{2}{*}{1} & 1 Branches & \multirow{2}{*}{FNN} & \multirow{2}{*}{4} & \multirow{2}{*}{[$6p$, $6p$, $p$]} & \multirow{2}{*}{tanh$\times$2 + linear} \\
		& & & 1 Trunk & & & & \\
		\midrule
		\multirow{2}{*}{\textit{Timescale 1}} & \multirow{2}{*}{1} & \multirow{2}{*}{1} & 2 Branches & \multirow{2}{*}{FNN} & \multirow{2}{*}{4} & \multirow{2}{*}{[$4p$, $4p$, $p$]} & \multirow{2}{*}{tanh$\times$2 + linear} \\
		& & & 1 Trunk & & & & \\
		\midrule
		\multirow{2}{*}{\textit{Timescale 2}} & \multirow{2}{*}{1} & \multirow{2}{*}{1} & 3 Branches & \multirow{2}{*}{FNN} & \multirow{2}{*}{12} & \multirow{2}{*}{[$4p$, $4p$, $p$]} & \multirow{2}{*}{tanh$\times$2 + linear} \\
		& & & 1 Trunk & & & & \\
		\midrule
		\multirow{6}{*}{\textit{Timescale 3}} & \multirow{6}{*}{1} & \multirow{2}{*}{1} & 3 Branches & \multirow{6}{*}{FNN} & \multirow{2}{*}{4} & \multirow{6}{*}{[$4p$, $4p$, $p$]} & \multirow{6}{*}{tanh$\times$2 + linear} \\
		& & & 1 Trunk & & & & \\
		& & \multirow{2}{*}{1} & 3 Branches & & \multirow{2}{*}{8} & & \\
		& & & 1 Trunk & & & & \\
		& & \multirow{2}{*}{1} & 3 Branches & & \multirow{2}{*}{12} & & \\
		& & & 1 Trunk & & & & \\
		\midrule
		\multirow{6}{*}{\textit{Timescale 4}} & \multirow{6}{*}{1} & \multirow{2}{*}{3} & 3 Branches & \multirow{6}{*}{FNN} & \multirow{2}{*}{4} & \multirow{6}{*}{[$4p$, $4p$, $p$]} & \multirow{6}{*}{tanh$\times$2 + linear} \\
		& & & 1 Trunk & & & & \\
		& & \multirow{2}{*}{3} & 3 Branches & & \multirow{2}{*}{8} & & \\
		& & & 1 Trunk & & & & \\
		& & \multirow{2}{*}{3} & 3 Branches & & \multirow{2}{*}{12} & & \\
		& & & 1 Trunk & & & & \\
		\bottomrule
	\end{tabular}
	\caption{\textit{CG-DeepONets architecture for 1-$\mathrm{D}$ shock case scenario.} $p$ is the dimension of the features embedding, FNN is the conventional feed-forward neural network, and $\sigma$ are the activation functions. Total parameters: 125$\,$460.}
	\label{table:archit_cg.1d}
\end{table}
\begin{table}[!htb]
	\centering
	\begin{tabular}{ccccccc}
		\toprule
		& \multirow{2}{*}{\# DeepONets} & \multicolumn{5}{c}{Single DeepONet} \\
		\cmidrule(lr){3-7}
		& & Sub-networks & Type & $p$ & Layers Width & $\sigma$ \\
		\midrule
		\multirow{2}{*}{\textit{Timescale 1}} & \multirow{2}{*}{1} & 1 Branches & \multirow{2}{*}{FNN} & \multirow{2}{*}{8} & \multirow{2}{*}{[24, 24, $p$]} & \multirow{2}{*}{tanh$\times$2 + linear} \\
		& & 1 Trunk & & & & \\
		\midrule
		\multirow{2}{*}{\textit{Timescale 2}} & \multirow{2}{*}{1} & 3 Branches & \multirow{2}{*}{FNN} & \multirow{2}{*}{16} & \multirow{2}{*}{[36, 36, $p$]} & \multirow{2}{*}{tanh$\times$2 + linear} \\
		& & 1 Trunk & & & & \\
		\midrule
		\multirow{2}{*}{\textit{Timescale 3}} & \multirow{2}{*}{3} & 3 Branches & \multirow{2}{*}{FNN} & \multirow{2}{*}{8} & \multirow{2}{*}{[24, 24, $p$]} & \multirow{2}{*}{tanh$\times$2 + linear} \\
		& & 1 Trunk & & & & \\
		\bottomrule
	\end{tabular}
	\caption{\textit{Neq-DeepONets architecture for 1-$\mathrm{D}$ shock case scenario.} $p$ is the dimension of the features embedding, FNN is the conventional feed-forward neural network, and $\sigma$ are the activation functions. Total parameters: 30$\,$325.}
	\label{table:archit_neq.1d}
\end{table}

\subsection{Computational framework}\label{ssec:shock.comp_frame}
To perform the numerical experiments presented in this work, three different software are used:
\begin{enumerate}[i.]
	\item \textsc{hegel} (High-fidElity tool for maGnEto-gasdynamics simuLations), a parallel multi-block structured fluid solver for LTE/NLTE plasmas written in modern object oriented \texttt{Fortran} 2008~\cite{Munafo_JComP_2020,Alberti_AIAA_2021,Alberti_JAP_2022}.
	\item \textsc{plato} (PLAsmas in Thermodynamic nOn-equilibrium), a physico-chemical library to evaluate thermodynamic, transport and optical properties as well as source terms due to NLTE collisional and radiative processes~\cite{Munafo_JComP_2020,Alberti_AIAA_2021,Alberti_JAP_2022}.
	\item PyCOMET (Physics-informed machine learning for scientific computing and operator discovery) is a TensorFlow-based~\cite{TF_2016} machine learning library that is used to construct neural operators and generic deep neural network (DNN)-based surrogates for scientific computing~\cite{Zanardi_AIAA_2022, Sharma_CRxiv_2022}. Previous approaches to integrating machine learning models into computational fluid dynamics (CFD) codes have included remote function calls from legacy \texttt{Fortran} codes to modern machine learning libraries, re-implementation of the full fluid solver in TensorFlow, or direct embedding of the network into the code. In this work, the approach used leverages the in-house \texttt{Fortran} and \texttt{C++} PyCOMET interfaces, which rely on CppFlow~\cite{cppflow_2019}, a \texttt{C++} wrapper of the TensorFlow \texttt{C} API. One significant benefit of this approach is its flexibility, as the interfaces can read and import any network or generic ML-based architectures into external codes without requiring complicated supplementary coding, and support both CPU and GPU operations.
\end{enumerate}
The \textsc{plato} library is responsible for performing the integration of the reactive step in \eqRef{eq:split.02} employed for evaluating the speedup in the one-dimensional shock case scenario (see the \ManuSecRef{One-dimensional shock case scenario}). The ODE integrator employed is the second-order backward differentiation formula (BDF-2) from the LSODE (Livermore Solver for Ordinary Differential Equations) library~\cite{Radhakrishnan_LSODE_1993}, with an absolute tolerance and a relative tolerance set to $10^{-9}$ and $10^{-6}$, respectively.

%% file: sections/extras/adapt_algo.tex
\begin{algorithm}[!htb]
	\relsize{-1}
	\caption{Adaptive inference}\label{alg:adapt}
	\vskip 1mm
	\Input{Initial conditions matrix and times instants vector
		\begin{equation*}
			X=\left[
			\hspace{1mm}
			\left\{\left(T^i,\rho^i,\boldsymbol{Y}^i_{{\text{O}_2}_{0}}\right)\right\}_{i=1}^n
			\in\mathbb{R}^{n\times\left(2+\mathcal{N_G}\right)},
			\hspace{1mm}
			\left\{t^i\right\}_{i=1}^n\in\mathbb{R}^{n}
			\hspace{1mm}
			\right]
		\end{equation*}
	}
	\Output{Mass fractions matrix
		\begin{equation*}
			Y=\left\{\widehat{\bm{Y}}^i\right\}_{i=1}^n
			\in\mathbb{R}^{n\times\left(1+\mathcal{N_G}\right)}
		\end{equation*}
	}
	\algDefine $TS$ = Number of timescales considered
	\vskip 0mm
	\algDefine $\delta_{tol}$ = Underpredicted non-equilibrium tolerance metric
	\vskip 2mm
	\textbf{Step 1}: Evaluate underpredicted non-equilibrium metric\\
	\For{$ts=1,\dots,TS-1$}{
		\algDefine $\mathcal{N_G}_{ts}$ = Number of groups in \textit{Timescale} $ts$\\
		\If{$ts=1$}
		{
			Compute $\bm{\delta}^{\left(1,1,1\right)}\in\mathbb{R}^{n}$ with Neq-DeepONet$^{\left(1,1\right)}$
		}
		\Else
		{
			\If{$ts\ne TS-1$ \algAnd all$\,\left(\left\{\delta_i^{\left(ts-1,\cdot,P\right)}\le\delta_{tol}\right\}_{i=1}^n\right)
				\hspace{1.5mm}\forall\hspace{1mm}P=1,\dots,\mathcal{N_G}_{ts-1}$}
			{
				\algBreak
			}
			\Else
			{
				\For{$P=1,\dots,\mathcal{N_G}_{ts-1}$}{
					\algDefine $\mathcal{N}_p$ = Number of micro-groups $p$ contained in macro-group $P$, $\mathcal{I}_p\subset\mathcal{I}_P$\\
					\If{any$\,\left(\left\{\delta_i^{\left(ts-1,\cdot,P\right)}>\delta_{tol}\right\}_{i=1}^n\right)$}
					{
						\begin{itemize}
							\item Mask out input data points for which $\left\{\delta_i^{\left(ts-1,\cdot,P\right)}\le\delta_{tol}\right\}_{i=1}^m$ with $m<n$\\
							\item Compute $\bm{\delta}^{\left(ts,\cdot,p\right)}\in\mathbb{R}^{n-m}
							\hspace{1.5mm}\forall\hspace{1mm}\mathcal{I}_p\subset\mathcal{I}_P$ with Neq-DeepONet$^{\left(ts,P\right)}$\\
							\item Assign $\bm{\delta}^{\left(ts,\cdot,p\right)}=\bm{0}\in\mathbb{R}^{m}
							\hspace{1.5mm}\forall\hspace{1mm}\mathcal{I}_p\subset\mathcal{I}_P$
						\end{itemize}
					}
					\Else
					{
						Assign $\bm{\delta}^{\left(ts,\cdot,p\right)}=\bm{0}\in\mathbb{R}^n
						\hspace{1.5mm}\forall\hspace{1mm}\mathcal{I}_p\subset\mathcal{I}_P$
					}
				}
			}
		}
	}
	\vskip 2mm
	\textbf{Step 2}: Employ computed underpredicted non-equilibrium metric to evaluate mass fractions\\
	\For{$ts=1,\dots,TS$}{
		\algDefine $\mathcal{N_G}_{ts}$ = Number of groups in \textit{Timescale} $ts$\\
		\If{$ts=1$}
		{
			Compute $\left\{\widehat{\bm{Y}}_{\text{O}},\,\widehat{\bm{Y}}_{\text{O}_2}^{(1,1,1)}\right\}\in\mathbb{R}^{n\times 2}$ with CG-DeepONet$^{\left(1,1\right)}$
		}
		\Else
		{
			\If{$ts\ne TS$ \algAnd  all$\,\left(\left\{\delta_i^{\left(ts-1,\cdot,P\right)}\le\delta_{tol}\right\}_{i=1}^n\right)
				\hspace{1.5mm}\forall\hspace{1mm}P=1,\dots,\mathcal{N_G}_{ts-1}$}
			{
				Reconstruct $\widehat{\bm{Y}}_{\text{O}_2}^{\left(TS,\cdot,p\right)}\in\mathbb{R}^n
				\hspace{1.5mm}\forall\hspace{1mm}\mathcal{I}_p\subset\mathcal{I}_P$ from $\widehat{\bm{Y}}_{\text{O}_2}^{(ts-1,\cdot,P)}\in\mathbb{R}^n\hspace{1.5mm}\forall\hspace{1mm}P=1,\dots,\mathcal{N_G}_{ts-1}$ by employing the Boltzmann distribution function
				\\\algBreak
			}
			\Else
			{
				\For{$P=1,\dots,\mathcal{N_G}_{ts-1}$}{
					\algDefine $\mathcal{N}_p$ = Number of micro-groups $p$ contained in macro-group $P$, $\mathcal{I}_p\subset\mathcal{I}_P$\\
					\If{any$\,\left(\left\{\delta_i^{\left(ts-1,\cdot,P\right)}>\delta_{tol}\right\}_{i=1}^n\right)$}
					{
						\begin{itemize}
							\item Mask out input data points for which $\left\{\delta_i^{\left(ts-1,\cdot,P\right)}\le\delta_{tol}\right\}_{i=1}^m$ with $m<n$\\
							\item Compute $\widehat{\bm{Y}}_{\text{O}_2}^{\left(ts,\cdot,p\right)}\in\mathbb{R}^{n-m}
							\hspace{1.5mm}\forall\hspace{1mm}\mathcal{I}_p\subset\mathcal{I}_P$ with CG-DeepONet$^{\left(ts,P\right)}$\\
							\item Reconstruct $\widehat{\bm{Y}}_{\text{O}_2}^{\left(ts,\cdot,p\right)}\in\mathbb{R}^m\hspace{1.5mm}\forall\hspace{1mm}\mathcal{I}_p\subset\mathcal{I}_P$ from $\widehat{\bm{Y}}_{\text{O}_2}^{(ts-1,\cdot,P)}\in\mathbb{R}^m$ by employing the Boltzmann\\distribution function
						\end{itemize}
					}
					\Else
					{
						Reconstruct $\widehat{\bm{Y}}_{\text{O}_2}^{\left(ts,\cdot,p\right)}\in\mathbb{R}^n\hspace{1.5mm}\forall\hspace{1mm}\mathcal{I}_p\subset\mathcal{I}_P$ from $\widehat{\bm{Y}}_{\text{O}_2}^{(ts-1,\cdot,P)}\in\mathbb{R}^n$ by employing the Boltzmann distribution function
					}
				}
			}
		}
	}
\vskip 1mm
\end{algorithm}